\documentclass[12pt,twoside,chap]{paper}
\usepackage{axodraw}
\usepackage{minitoc,cite,fancyhdr}
\usepackage{bbm} 
\usepackage{euscript,amssymb,latexsym} 
\usepackage{epsf,epsfig,graphics,subfigure,graphicx}
\usepackage{amsmath}
     
\usepackage{eepic}
                 
\makeatletter
\def\fmslash{\@ifnextchar[{\fmsl@sh}{\fmsl@sh[0mu]}}
\def\fmsl@sh[#1]#2{%
  \mathchoice
    {\@fmsl@sh\displaystyle{#1}{#2}}%
    {\@fmsl@sh\textstyle{#1}{#2}}%
    {\@fmsl@sh\scriptstyle{#1}{#2}}%
    {\@fmsl@sh\scriptscriptstyle{#1}{#2}}}
\def\@fmsl@sh#1#2#3{\m@th\ooalign{$\hfil#1\mkern#2/\hfil$\crcr$#1#3$}}
\makeatother
\global\arraycolsep=2pt 

\def\be{\begin{equation}}
\def\ee{\end{equation}}
\newcommand{\bea}{\begin{eqnarray}}
\newcommand{\eea}{\end{eqnarray}}

\makeatletter
\def\cleardoublepage{\clearpage\if@twoside \ifodd\c@page\else%
\hbox{}%
\thispagestyle{empty}%
\newpage%
\if@twocolumn\hbox{newpage}\fi\fi\fi}
\makeatother

\begin{document}
\title{A Positive-Weight Next-to-Leading-Order Monte Carlo Simulation of 
       Deep Inelastic Scattering and Higgs Boson Production via Vector Boson Fusion in Herwig++}

\author{\bf{\normalsize{Luca D'Errico}}
\\ 
\vspace{-0.3 cm} \it{\normalsize{Institut f\"ur Theoretische Physik,}}
\\ 
\vspace{-0.1 cm} \it{\normalsize{University of Karlsruhe, KIT, 76128, Germany;}}
\\
\vspace{-0.3 cm} \it{\normalsize{Institute of Particle Physics Phenomenology, Department of Physics,}}
\\ 
\vspace{-0.1 cm} \it{\normalsize{University of Durham, DH1 3LE, UK;}}
\\
\it{\normalsize{Email:} }
\sf{\normalsize{derrico@particle.uni-karlsruhe.de}}
\\
\\
\vspace{-0.1 cm} \bf{\normalsize{Peter Richardson}} 
\\ 
\vspace{-0.3 cm} \it{\normalsize{Institute of Particle Physics Phenomenology, Department of Physics,}}
\\
\vspace{-0.1 cm} \it{\normalsize{University of Durham, DH1 3LE, UK;}}
\\
\it{\normalsize{Email:} }
\sf{\normalsize{peter.richardson@durham.ac.uk}}
}
\maketitle
\vspace{-15cm}
\begin{flushright}
KA-TP-12-2011\\
SFB/CPP-11-29\\
MCNET-11-14 \\
DCPT/11/66\\
IPPP/11/33 
\end{flushright}
\vspace{13.5cm}
\begin{abstract}
The positive weight next-to-leading-order matching formalism (POWHEG) is applied to 
Deep Inelastic Scattering~(DIS) and the related 
Higgs boson production via vector-boson fusion process in the \textsf{Herwig++}
Monte Carlo event generator. This scheme combines parton shower simulation
and next-to-leading-order calculation in a consistent way which only produces positive 
weight events. The simulation contains a full implementation of the truncated shower
required to correctly model soft emissions in an angular-ordered parton shower. 
\end{abstract}
\thispagestyle{empty}
\newpage
\thispagestyle{plain}
\setcounter{page}{1}
---------------------------------------------------------------------------------------------------------
\tableofcontents
---------------------------------------------------------------------------------------------------------
\section{Introduction}

The Large Hadron Collider (LHC) at CERN is designed to elucidate the nature of electroweak
symmetry breaking in the Standard Model\cite{Englert:1964et, Higgs:1964pj, Guralnik:1964eu, Kibble:1967sv}, and in particular
discover the Higgs boson. Once the Higgs boson has been observed 
and its mass determined, it will be
crucial to measure the way it couples to gauge bosons and 
fermions~\cite{Zeppenfeld:2000td, Belyaev:2002ua}. 
The most promising processes in which these couplings of the Higgs boson can be measured
are gluon-gluon and vector-boson fusion. 
The former consists of a gluon-gluon partonic collision which produces the Higgs boson
via a virtual top quark loop~\cite{Georgi:1977gs}.
It has the largest cross section for Higgs boson masses less than $\sim 1$\,TeV and will
be important for the measurement of the Higgs boson coupling to the top quark. 

Higgs boson production via vector-boson fusion (VBF) is a process in which 
two incoming fermions each radiate a $W^\pm$ or $Z^0$ boson which then combine
to produce the Higgs boson.
VBF is expected to play a fundamental r\^ole in the measurement of the Higgs boson
couplings to gauge bosons and fermions, because it allows for independent
observation in different channels:
$H \to \tau\tau$~\cite{Rainwater:1998kj, Plehn:1999xi}, $H\to WW$~\cite{Rainwater:1999sd, Kauer:2000hi}, 
$H\to \gamma\gamma$~\cite{Rainwater:1997dg} and $H\to$invisible~\cite{Eboli:2000ze, Cavalli:2002vs}.

In order to calculate the Higgs boson coupling constants with sufficient accuracy, 
next-to-leading-order~(NLO) QCD corrections for the VBF process must be included. These
corrections have been known for some time~\cite{Han:1991ia} and are relatively small
with $K$-factors around $1.05$ to $1.1$. 
At next-to-leading-order, the theoretical prediction of the Standard Model
production cross sections have an error of less than $10\%$. This accuracy
is sufficient to compare predictions with upcoming LHC measurements, 
which will be performed with a statistical accuracy on the product of the production cross section and
decay branching ratio reaching $5$ to $10 \%$~\cite{Zeppenfeld:2000td, Belyaev:2002ua}. 
The theoretical uncertainties for the VBF process therefore do not significantly compromise
the precision of the coupling constant measurements. 
This makes the VBF process more attractive than Higgs boson production via gluon fusion, which has a K-factor larger 
than~$2$ and for which the uncertainties remain between $10-20 \%$ even after
the inclusion of next-to-next-to-leading-order
corrections~\cite{Djouadi:1991tka, Spira:1995rr, Dawson:1990zj, Giele:2002hx, Catani:2001ic, Harlander:2001is, Harlander:2002wh, Anastasiou:2002yz, Ravindran:2003um}.
Nevertheless, stringent cuts are necessary to distinguish the VBF Higgs boson signal
from the backgrounds. In particular, a veto on additional activity in the events,
the central-jet veto, is often imposed to reduce the backgrounds.

In order to study the effects of these cuts we must rely on
Monte Carlo event generators which combine, usually leading-order~(LO), matrix elements
with parton showers and hadronization models to provide a fully exclusive simulation
of an event. Traditionally leading-order matrix elements have been used in these
simulations together with the parton shower approximation which simulates soft and
collinear emission. However, in recent years a number of different approaches have been
developed to improve the simulation of high transverse momentum, $p_{T}$, 
radiation\footnote{See Ref.\,\cite{Buckley:2011ms} for a recent review of the older
techniques~\cite{Sjostrand:1986hx, Bengtsson:1987rw, Norrbin:2000uu, Miu:1998ju,Corcella:2000bw,Corcella:2002jc,Seymour:1991xa,Seymour:1994ti,Corcella:1998rs,Corcella:1999gs,Seymour:1994df,Seymour:1994we,Gieseke:2003hm, Gieseke:2004af, Hamilton:2006ms, Gieseke:2006ga, Bahr:2008tx, Gieseke:2011na}
and techniques for improving the simulation of multiple hard QCD
radiation~\cite{Catani:2001cc, Krauss:2002up, Lonnblad:2001iq, Schalicke:2005nv, Krauss:2005re, Lavesson:2005xu, Mrenna:2003if, Mangano:2002ea, Alwall:2007fs,Hoeche:2009rj,Hamilton:2009ne}.}.

A number of approaches has been developed to provide a description
of hardest emission together with a cross section which is accurate to
next-to-leading-order. In the
approach of Frixione and Webber (MC@NLO)~\cite{Frixione:2002ik,Frixione:2010wd},
the parton shower approximation is subtracted from the exact next-to-leading-order
calculation. This was the first successful systematic scheme for matching
next-to-leading-order calculations and parton showers and has been applied to
many different processes~\cite{Frixione:2005vw,Frixione:2007zp,Frixione:2008yi,LatundeDada:2007jg, LatundeDada:2009rr,Papaefstathiou:2009sr,Torrielli:2010aw,Frixione:2010ra}. 
However, this method has two drawbacks: it generates weights which are not
positive definite and is implemented in a way which is fundamentally dependent
on the details of the parton shower algorithm.

These problems have been addressed with a new matching algorithm proposed by Nason, POWHEG~(POsitive Weight Hardest Emission
Generator)~\cite{Nason:2004rx, Frixione:2007vw}, which achieves
the same aims as MC@NLO but produces only positive weight events. Although it is independent of the generator
with which it is implemented, it requires the shower to have a particular structure:
a \textit {truncated shower} simulating wide angle soft emission; followed 
by the highest transverse momentum~($p_{T}$) 
parton emission; followed again by a \textit{vetoed shower} simulating softer radiation. The 
highest $p_{T}$ emission is generated separately using a Sudakov form factor containing
the real emission piece of the differential cross section. The \textit{truncated shower}
produces radiation at a higher scale~(in the evolution variable of the parton shower),
and the \textit{vetoed shower} at a lower
scale, than the one at which the hardest emission is generated. The POWHEG method
has been applied to wide range of processes~\cite{Nason:2006hfa,Frixione:2007nw,LatundeDada:2006gx,Alioli:2008gx, Hamilton:2008pd,Alioli:2008tz, Hamilton:2009za,Alioli:2009je,Nason:2009ai,Hoche:2010pf,Alioli:2010xd,Re:2010jg,Re:2010bp,Alioli:2010qp,Alioli:2010xa,Hamilton:2010mb,Oleari:2010nx,Oleari:2011ey,Kardos:2011qa,Melia:2011gk}\footnote{
There has also been some work combining either many NLO matrix elements~\cite{Lavesson:2008ah} or the NLO matrix elements with subsequent emissions matched to leading-order matrix elements~\cite{Hamilton:2010wh,Hoche:2010kg} with the parton shower.}.

For the VBF process, as the central-jet veto 
is sensitive to additional radiation in the event,
it is important to have an accurate simulation which gives both the NLO
cross section for the process and provides an accurate simulation
of additional QCD radiation. In this work we will describe the simulation
of this process in \textsf{Herwig++}~\cite{Bahr:2008pv,Gieseke:2011na} using the POWHEG approach.

As we develop new simulations it is important to validate them
using experimental data wherever possible. While obviously 
there is no data on the vector boson fusion process,
Deep Inelastic Scattering~(DIS) has many of
the same features, in particular in both processes it is important 
that the parton shower algorithm preserves the momentum of the space-like
vector bosons. In addition the wealth of data from HERA makes deep inelastic
scattering important for the tuning of the phenomenological parameters in 
Monte Carlo event generators. There are existing simulations of the
VBF~\cite{Nason:2009ai} and DIS processes~\cite{Hoche:2010pf} in the POWHEG approach. However, due to different treatment of this class of processes in the angular-ordered
parton shower in \textsf{Herwig++}~(the different kinematical reconstruction of these processes,
preserving the virtuality of the $t$-channel gauge bosons, and generation of the truncated shower)
and the experimental importance of this processes 
it is important to have a range of different simulations with different approaches. In addition our factorized approach
makes the extension to other colour-singlet vector-boson fusion processes
simple, requiring only a calculation
of the leading-order matrix element.

The rest of the paper is organized as follows. In Sect.~\ref{sec1} we recap
the POWHEG formulae of main interest for our description. The calculation of
the leading-order kinematics with NLO accuracy in the POWHEG approach will be
discussed in Sect.~\ref{sec2}. A brief description of the generation of the
hardest emission within \textsf{Herwig++} will be outlined in Sect.~\ref{sec3} and
in Sect.~\ref{sec3second} we give details of the implementation of truncated and
vetoed showers in the program.  Our results will be described in Sect.~\ref{sec4}. 
We finally present our conclusions in Sect.~\ref{sec5}. 

\section{The POWHEG method} \label{sec1}

In this section we introduce the details of the POWHEG algorithm. 
The NLO differential cross section for a given N-body process, 
can be written within the POWHEG approach as
\begin{equation}
\mathrm{d}\sigma=\bar{B}(\Phi_{B})\mathrm{d}\Phi_{B}
\left[\Delta_{R}(0)+\frac{R(\Phi_{B},\Phi_{R})}{B(\Phi_{B})}
\Delta_{R}(k_{T}(\Phi_{B},\Phi_{R}))\mathrm{d}\Phi_{R}\right]\text{,}\label{lo_scs}
\end{equation}
where $\bar{B}(\Phi_{B})$ is defined as
\begin{equation}
\bar{B}(\Phi_{B})=B(\Phi_{B})+V(\Phi_{B})
+\int \left[ R(\Phi_{B},\Phi_{R})-\sum_{I}D_{I}(\Phi_{B},\Phi_{R})\right]
\mathrm{d}\Phi_{R}\text{,}
\end{equation}
$B(\Phi_{B})$ is the leading-order contribution, $\Phi_{B}$ the N-body phase-space
variables of the LO process, $V(\Phi_{B})$ a finite contribution including
unresolvable, real emission and virtual loop pieces, $\Phi_{R}$ are the
radiative variables describing the phase space for the emission of an extra parton,
$R(\Phi_{B},\Phi_{R})$ is the matrix element including the radiation of an
additional parton multiplied by the relevant parton flux factors and
$D_{I}(\Phi_{B},\Phi_{R})$ are the counter terms regulating the singularities
in $R(\Phi_{B},\Phi_{R})$. 
The modified Sudakov form factor is defined in terms of $R(\Phi_{B},\Phi_{R})$ as
\begin{equation}
\Delta_{R}(p_{T})=\exp \left [-\int \mathrm{d}\Phi_{R}
\frac{R(\Phi_{B},\Phi_{R})}{B(\Phi_{B})}
\theta(k_{T}(\Phi_{B},\Phi_{R})-p_{T}) \right]\text{,} \label{mod_sud}
\end{equation}
where $k_{T}(\Phi_{B},\Phi_{R})$ is equal to the transverse momentum of the
emitted parton in the soft and collinear limits.

The POWHEG formalism requires that the N-body configuration
is generated according to $\bar{B}(\Phi_{B})$. The hardest emission in
the event is then generated using the Sudakov form factor given in
Eqn.\,\ref{mod_sud}.
As $\bar{B}(\Phi_{B})$ is simply the next-to-leading-order differential cross
section integrated over the radiative variables, it is naturally positive, 
and therefore leads to the absence of events with negative weights.

If the parton shower simulation is ordered in transverse momentum we can
simulate the process by first generating the hardest emission and then evolving
the process using the parton shower from the $N+1$ parton final state forbidding any
emissions with transverse momentum above that of the hardest one.
However, for shower algorithms which are ordered in other variables, for
example angular ordering in \textsf{Herwig++}, the hardest $p_T$ emission is not
generated as the first emission in the parton shower.
So the shower must be reorganized into a truncated shower
which describes soft emissions, at higher evolution scales than the 
highest $p_T$ emission,
together with vetoed showers which describe emissions at lower
evolution scales which are constrained to be softer than the hardest
emission~\cite{Nason:2004rx,Frixione:2007vw}.

We implement the POWHEG algorithm according to the following procedure:
\begin{itemize}
\item generate an event according to Eqn.\,\ref{lo_scs};
\item directly hadronize the small fraction of non-radiative events;
\item map the radiative variables parameterizing the emission into the evolution
 scale, momentum fraction and azimuthal angle, $(\tilde{q}_{h},z_{h},\phi_{h})$,
 from which the parton shower would reconstruct identical momenta;
\item consider the initial $N$-body configuration generated from $\bar{B}(\Phi_{B})$
 and evolve the parton emitting the extra radiation from the default starting scale
 down to $\tilde{q}_{h}$ using the truncated shower;
\item insert a branching with parameters $(\tilde{q}_{h},z_{h},\phi_{h})$ into the
 shower when the evolution scale reaches $\tilde{q}_{h}$;
\item generate $p_{T}$ vetoed showers from all external legs. 
\end{itemize}
This simple approach allows us to correctly generate wide-angle
soft radiation using the truncated shower which was absent in some earlier
simulations.

\section{Calculation of \boldmath{$\bar{B}(\Phi_{B})$}}\label{sec2}

\begin{figure}[t]
\centering
\includegraphics[width=\textwidth]{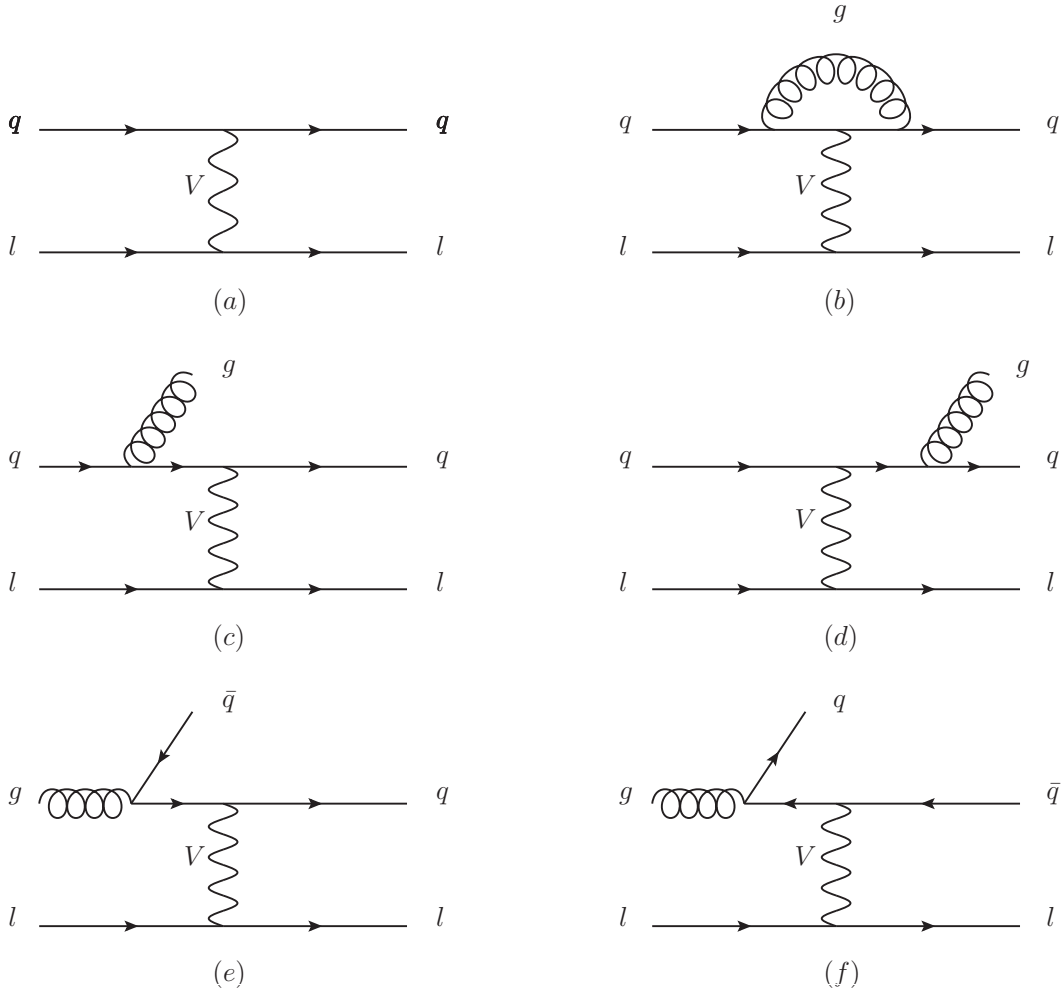}
\caption{Feynman diagrams contributing to deep inelastic scattering
         at $\mathcal{O}(\alpha_{s})$: leading order (a), virtual (b) and real emission (c-f)
         corrections.}
\label{dis_graph}
\end{figure}
\begin{figure}[t]
\centering
\includegraphics[width=\textwidth]{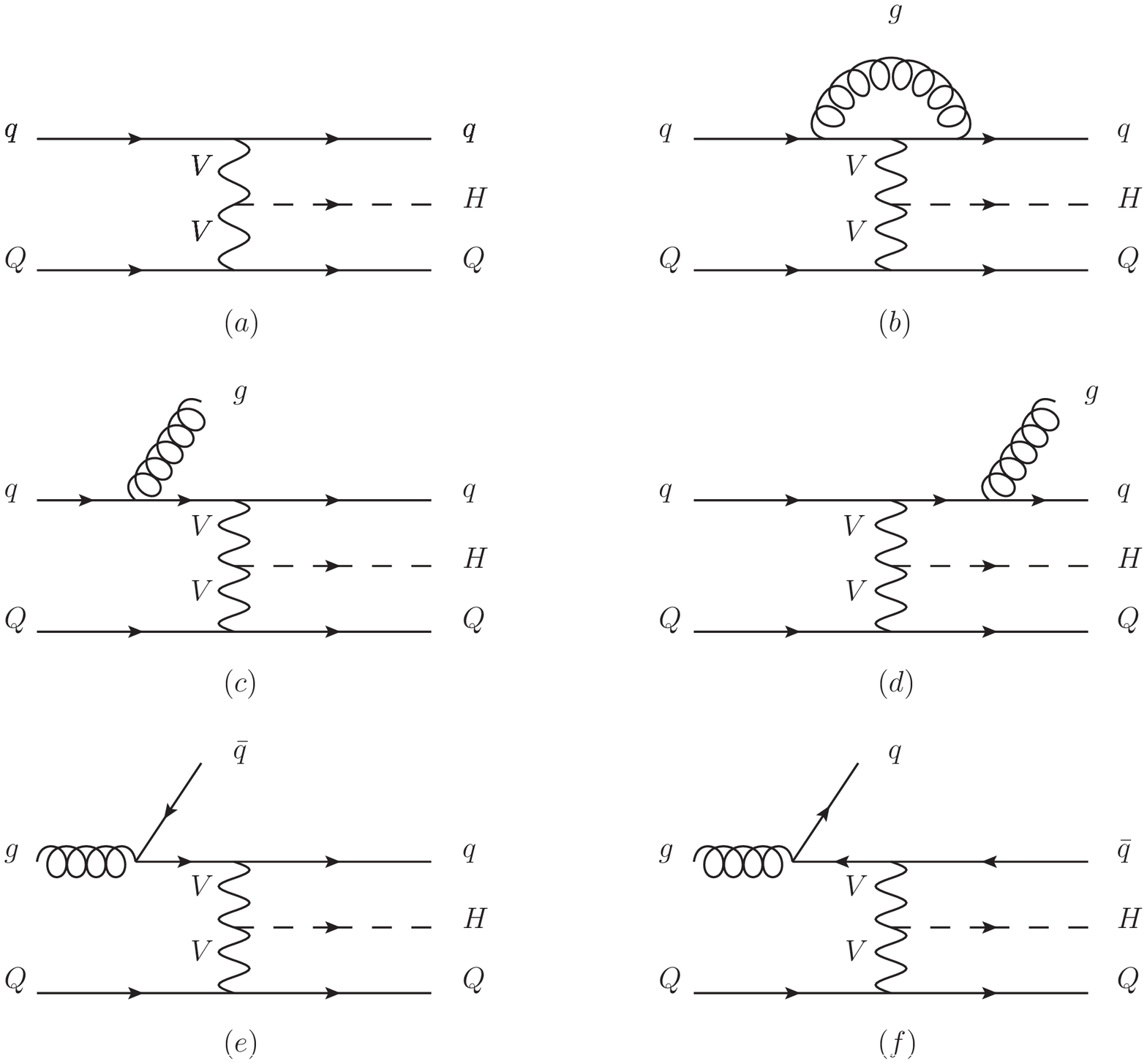}
\caption{Feynman diagrams contributing to Higgs boson production via weak-boson fusion
         at $\mathcal{O}(\alpha_{s})$: leading order (a), virtual (b) and real emission (c-f)
         corrections. As matter of simplicity, we just show radiative corrections to the upper
         line in the gluon emission (c-d) and gluon initiated processes (e-f). In this case $V=W^\pm$ or $Z^0$.}\label{vbf_graph}
\end{figure}

At leading order DIS is described by the Feynman diagram shown in Fig.\,\ref{dis_graph}a,
together with appropriate crossings of the quark line. The leading-order
diagram for the VBF process is shown in Fig.\,\ref{vbf_graph}a, together with 
appropriate crossings of the quark lines.
In principle other contributions to the VBF process should be considered: 
diagrams with the exchange of identical outgoing quarks,
the quark annihilation processes $\bar{q}q \to Z^{*} \to ZH$
and $\bar{q}q \to W^{\pm *} \to W^\pm H$ with hadronic decays of the vector bosons.
However, colour suppression
and the large momentum transfer in the weak-boson propagators make the contribution
coming from these additional processes negligible in the phase-space regions where
VBF can be observed experimentally, {\it i.e.} with widely separated quark jets
of very large invariant mass \cite{Figy:2003nv}. 

At $\mathcal{O}(\alpha_{s})$, the contributions coming from amplitudes in
which the gluon is attached to both upper and lower quark lines in the VBF process
vanish because the weak boson has no colour charge. The Feynman graphs contributing
are therefore only the ones shown in Figs.\,\ref{vbf_graph}b-\ref{vbf_graph}f 
where for simplicity, we just show radiation from the upper quark line. 

The corrections to the DIS and VBF processes are therefore the same provided that
we take into account the corrections to both of the quark lines in the VBF process.
In this section, we show the analytical contributions
to the next-to-leading-order differential cross section for DIS and VBF. 
Collecting the real emission cross section, described in Sect.~\ref{subsec1}, 
virtual and collinear contributions, briefly discussed in Sect.~\ref{subsec2},
$\bar{B}(\Phi_{B})$ can be calculated. 
We then discuss how it is sampled within \textsf{Herwig++} in Sect.~\ref{subsec3}. 

\subsection{Real emission contribution}\label{subsec1}

The tree-level corrections to $e^{+}e^{-}$ annihilation to hadrons can be
written in a form in which QCD and electroweak parts exactly
factorize~\cite{Kleiss:1986re}.
This method was later generalized to any process in which the
lowest-order diagrams
contain a single quark line attached to a single electroweak 
gauge boson~\cite{Seymour:1994we}.
We adopt this approach for the calculation of the real corrections to DIS and VBF,
based on the calculation of the correction to DIS in 
Refs.~\cite{Seymour:1994we,Seymour:1994ti}.
\begin{figure}
\centering
\includegraphics[width=\textwidth]{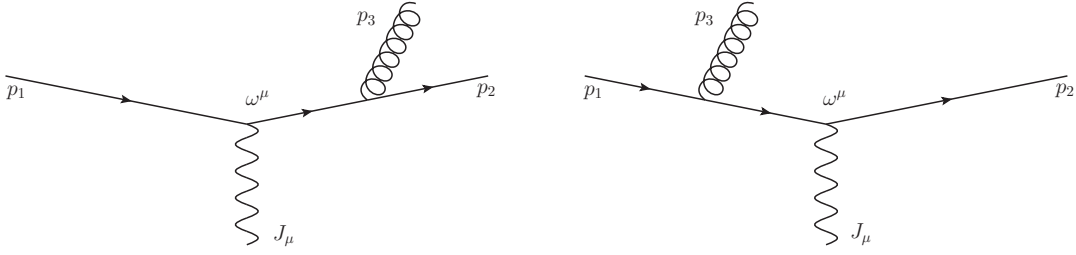}
\caption{QCD Compton scattering, in which a quark interacts with an arbitrary 
         current $J_{\mu}$ via boson-quark coupling $\omega^{\mu}$.}\label{feym_graph}
\end{figure}

Consider the QCD Compton process, shown in Fig.\,\ref{feym_graph}, 
where a quark $q$ with momentum $p_{1}$ and a fraction $x_{B}$ of the
incoming hadron momentum, interacts with a current $J_{\mu}$ and boson-parton
coupling $\omega^{\mu}$, and scatters into an outgoing quark $q^{\prime}$ with
momentum $p_{2}$ and a gluon with momentum $p_{3}$
\begin{equation}
\text{QCDC}: q(p_{1})+X \rightarrow q^{\prime}(p_{2})+g(p_{3})+X^{\prime}\text{.}
\end{equation}
It is simplest to work in the \textit{Breit frame}, in which the incoming parton
for the leading-order process has four-momentum $q_1=\frac{Q}{2}(1;0,0,1)$, the
exchanged boson has four-momentum $q=(0;0,0,-Q)$ and the scattered quark
has four-momentum $q_2=\frac{Q}2(1;0,0,-1)$.
In this frame the four-momenta of the real emission process are
\begin{subequations}
\begin{eqnarray}
p_1 &=& \frac{Q}2(-x_1;0,0,-x_1);\\
p_2 &=& \frac{Q}2(\sqrt{x^2_2+x_\perp^2};\phantom{-}x_\perp\cos\phi,
                                        \phantom{-}x_\perp\sin\phi,-x_2);\\
p_3 &=& \frac{Q}2(\sqrt{x^2_3+x_\perp^2};          -x_\perp\cos\phi,
                                                  -x_\perp\sin\phi,-x_3);
\end{eqnarray}
\end{subequations}
where the transfered momentum $q=(0;0,0,-Q)=p_{2}+p_{3}-p_{1}$ and
\begin{equation}
x_{i} =  \frac{2p_{i}\cdot q}{q \cdot q}\text{.}\label{xi}
\end{equation}
Momentum conservation requires that $x_{3}=2+x_{1}-x_{2}$ and
\begin{equation}
x_\perp^2 = \frac{(x_3^2-x_1^2-x_2^2)^2}{4x_1^2}-x^2_2.
\end{equation}

In terms of these variables the cross section for the real emission process is
\begin{equation}
{\rm d}\sigma_{\rm NLO} = \frac1{4(2\pi)^2}\frac{{\rm d}\phi}{2\pi}
\frac{{\rm d}x_1{\rm d}x_2}{-x_1^3} \frac{-x_1x_Bf(-x_1x_B,Q^2)}{x_Bf(x_B,Q^2)} \frac{Q^2|\mathcal{M}_{\rm QCDC}|^2}{|\mathcal{M}_{\rm LO}(q_1,q_2)|^2}  {\rm d}\sigma_2,
\end{equation}
where $x_B$ is the momentum fraction of the quark in the leading-order process,
$|\mathcal{M}_{\rm QCDC}|^2$ and $|\mathcal{M}_{\rm LO}(q_1,q_2)|^2$ are the spin and colour
averaged matrix elements squared for the real emission and leading-order
processes respectively.

  Using the gauge introduced by the CALKUL collaboration~\cite{DeCausmaecker:1981bg},
  the matrix element for the real
  emission process is~\cite{Seymour:1994we,Seymour:1994ti}
\begin{equation}
|\mathcal{M}_{\rm QCDC}|^2 = -\frac{8\pi\alpha_SC_F}{(1+x_1)(1-x_2)Q^2}
\left(x_1^2+(x_2^2+x_\perp^2)R_2\right)|\mathcal{M}_{\rm LO}(q_1,q_2)|^2,
\end{equation}
where
\begin{equation}
R_2 \equiv \frac{x_2^2}{x^2_2+x^2_\perp} \frac{|\mathcal{M}_{\rm LO}(\bar{r}_1,r_2)|^2}{|\mathcal{M}_{\rm LO}(q_1,q_2)|^2},
\end{equation}
with $r_2=\frac{p_2}{x_2}$ and $\bar{r}_1=r_2-q$.

The integration of the phase space is simpler if we use the variables:
\begin{subequations}
\begin{eqnarray}
x_1 &=& -\frac1{x_p};\\
x_2 &=& 1-\frac{1-z_p}{x_p};
\end{eqnarray}
\end{subequations}
so that the phase-space limits become $x_B<x_p<1$ and $0<z_p<1$. In terms of
these variables
\begin{equation}
x_{\perp}^{2}=\frac{4(1-x_{p})(1-z_{p})z_{p}}{x_{p}}\text{.}
\end{equation}

The cross section for the real emission process becomes
\begin{equation}
{\rm d}\sigma_{\rm NLO} = \frac{\alpha_S C_F}{2\pi}\frac{{\rm d}\phi}{2\pi}
\frac{\frac{x_B}{x_p}f(\frac{x_B}{x_p},Q^2)}{x_Bf(x_B,Q^2)} 
\frac{{\rm d}x_p{\rm d}z_p}{(1-x_p)(1-z_p)}\left(1+x_p^2(x_2^2+x_\perp^2)R_2\right)
{\rm d}\sigma_2.
\end{equation}
This allows us to treat the QCD Compton process as a correction to the leading-order
quark scattering process.

The boson-gluon fusion process,
\begin{equation} 
\text{BGF}: g(p_{1})+X \rightarrow q^{\prime}(p_{2})+\bar{q}(p_{3})+X^{\prime}\text{,}
\end{equation}
can be treated in a similar way.
In this case the spin and colour averaged matrix element squared is
\begin{equation}
|\mathcal{M}_{\rm BGF}|^2 = \frac{8\pi\alpha_ST_R}{(1-x_2)(1-x_3)Q^2}
\left((x_2^2+x_\perp^2)R_2+(x_3^2+x_\perp^2)R_3\right),
\end{equation}
where
\begin{eqnarray}
R_{3}=\frac{x_{3}^{2}}{x_{3}^{2}+x_{\perp}^{2}}
\frac{|\mathcal{M}_{\rm LO}(r_{3},r_3+q)|^ {2}}{|\mathcal{M}_{\rm LO}(q_{1},q_{2})|^ {2}}\text{,}
\end{eqnarray}
with $r_3=-p_3/x_3$.

Using the same change of variables as before the differential cross section is
\begin{equation}
{\rm d}\sigma_{\rm NLO} = \frac{T_R\alpha_S}{2\pi}\frac{{\rm d}\phi}{2\pi}
\frac{{\rm d}x_p{\rm d}z_p}{z_p(1-z_p)} \frac{\frac{x_B}{x_p}f\left(\frac{x_B}{x_p},Q^2\right)}{x_Bf(x_B,Q^2)}
\left(x_P^2(x_2^2+x_\perp^2)R_2+x_P^2(x_3^2+x_\perp^2)R_3\right).\label{bgf_cs}
\end{equation}
 So far, our result gives the calculation of the BGF
 cross section, without any distinction between quark and antiquark scattering. If we
 want to view Eqn.\,\ref{bgf_cs} as a correction to a given lowest-order process,
 partons and antipartons should be treated equivalently. As the
 $z_{p}=1$ singularity is associated with configurations that become collinear to the
 lowest-order quark scattering process, while the $z_{p}=0$ singularity is associated
 with the antiquark scattering process, we can separate
\begin{equation}
\frac{1}{z_{p}(1-z_{p})}=\frac{1}{z_{p}}+\frac{1}{1-z_{p}}
\end{equation}
and rewrite the cross section as
\begin{equation}
{\rm d}\sigma_{\rm NLO} = \frac{T_R\alpha_S}{2\pi}\frac{{\rm d}\phi}{2\pi}
\frac{{\rm d}x_p{\rm d}z_p}{(1-z_p)} \frac{\frac{x_B}{x_p}f(\frac{x_B}{x_p})}{x_Bf(x_B)}
\left(x_P^2(x_2^2+x_\perp^2)R_2+x_P^2(x_3^2+x_\perp^2)R_3\right),
\end{equation}
with the corresponding $\frac1{z_p}$ term giving a correction
to the antiquark scattering process~\cite{Seymour:1994we,Seymour:1994ti}.

Using these results we can rewrite the real emission corrections as
\begin{equation}
{\rm d}\sigma_R \equiv R(\Phi_{B},\Phi_{R})\mathrm{d}\Phi_{B}\mathrm{d}\Phi_{R},
\end{equation}
where
\begin{equation}
R(\Phi_{B},\Phi_{R}) = \sum_{I}R_{I}
= B(\Phi_{B})\sum_{I}\frac{\mathcal{C}_{I}
\alpha_{s}(\mu_{R})}{2\pi}\mathcal{A}_{I}\text{,}
\end{equation}
with $I\in\left\{{\rm QCDC},{\rm BGF}\right\}$ and
\begin{subequations}
\begin{eqnarray}
\mathcal{C}_{\mathrm{QCDC}} &=& C_{F}\text{,}\\
\mathcal{C}_{\mathrm{BGF}} &=& T_{R}\text{,}\\
\mathcal{A}_{\mathrm{QCDC}} &=& \frac{\frac{x_{B}}{x_{p}} f_{q}(\frac{x_{B}}{x_{p}},Q^{2})}
 {x_{B} f_{q}(x_{B},Q^{2})} 
\frac{1}{(1-x_{p}) (1-z_{p})}
\left(1 + x_p^2(x_{p}^{2}+x_{\perp}^{2}))R_{2}\right)\text{,}\\
\mathcal{A}_{\mathrm{BGF}} &=& \frac{ 
\frac{x_{B}}{x_{p}}f_{g}(\frac{x_{B}}{x_{p}},Q^{2})}
{x_{B} f_{q}({x_{B},Q^{2})}}
\frac{1}{(1-z_{p})}
\left(x_P^2(x_2^2+x_\perp^2)R_2+x_P^2(x_3^2+x_\perp^2)R_3\right)\text{.}
\end{eqnarray}
\end{subequations}
The radiative phase-space element is
\begin{equation}
\mathrm{d}\Phi_{R} = \frac{1}{2\pi}\mathrm{d}x_{p}\mathrm{d}z_{p}\mathrm{d}\phi\text{.}
\end{equation}

The singularities in $R(\Phi_{B},\Phi_{R})$ are cancelled by subtracting 
\begin{equation}
D_{I}=\frac{\mathcal{C}_{I}\alpha_{s}(\mu_{R})}{2\pi}\mathcal{D}_{I}\text{,}
\end{equation}
where $\mathcal{D}_{I}$ are the Catani-Seymour dipoles \cite{Catani:1996vz}:
\begin{subequations}
\begin{eqnarray}
\mathcal{D}_{\mathrm{QCDC}}&=&\frac{\frac{x_{B}}{x_{p}}f_{g}(\frac{x_{B}}{x_{p}},Q^{2})}{x_{B} f_{q}(x_{B},Q^{2})}\frac{x_p^{2}+z_p^{2}}{(1-x_p)(1-z_p)}\text{;}\\
\mathcal{D}_{\mathrm{BGF}}&=&\frac{\frac{x_{B}}{x_{p}}f_{g}(\frac{x_{B}}{x_{p}},Q^{2})}{x_{B} f_{q}(x_{B},Q^{2})}\frac{x_p^{2}+(1-x_p)^{2}}{1-z_p}\text{.}
\end{eqnarray}
\end{subequations}

The contribution of the real emission processes to $\bar{B}$ is therefore
\begin{equation}
B(\Phi_{B})\mathrm{d}\Phi_{B} \sum_{I\in\left\{{\rm QCDC},{\rm BGF}\right\}}\frac{\mathcal{C}_{I}
\alpha_{s}(\mu_{R})}{2\pi}\left(\mathcal{A}_{I}-\mathcal{D}_I\right)\mathrm{d}\Phi_{R}\text{.}
\end{equation}

While these expressions are sufficient for the calculation of the real correction, in the
case of DIS the leading-order matrix elements are simple enough that
$R_{2,3}$ can be calculated analytically and integrated over the azimuthal angle, $\phi$
and $z_p$ to simplify the numerical integration of $\bar{B}$.

For DIS \cite{Seymour:1994we,Seymour:1994ti},
\begin{equation}
R_2 = \frac{\cos^2\theta_2+\mathcal{A}\cos\theta_2\left(\ell-\sqrt{\ell^2-1}\sin\theta_2\cos\phi\right)+\left(\ell-\sqrt{\ell^2-1}\sin\theta_2\cos\phi\right)^2}{1+\mathcal{A}\ell+\ell^2},
\end{equation}
where $\cos\theta_2=\frac{x_2    }{\sqrt{x^2_2+x_\perp^2}}$,
      $\sin\theta_3=\frac{x_\perp}{\sqrt{x^2_2+x_\perp^2}}$, $\ell=\frac2{y_B}-1$, $y_B=\frac{q\cdot q_1}{p_\ell\cdot q_1}$ and $p_\ell$ is
the four-momentum of the incoming lepton.
$\mathcal{A}$ is related to the couplings of the fermions to the
exchanged vector bosons.
For the charged current process $\mathcal{A}=2$, whereas for the 
neutral current process
\begin{equation}
\mathcal{A} = \frac{4 r C_{A,\ell} C_{A,q} \left(Q_\ell Q_q+2rC_{V,\ell}C_{V,q} \right)}
{\left(Q_\ell^2Q_q^2+2Q_\ell Q_q rC_{V,\ell}C_{V,q}
+r^2\left(C^2_{V,\ell}+C^2_{A,\ell}\right)\left(C^2_{V,q   }+C^2_{A,q   }\right)\right)},
\end{equation}
with $r=\frac{Q^2}{(Q^2+m^2_Z)}$ and
\begin{subequations}
\begin{eqnarray}
C_{V,i} &=& \frac1{\sin\theta_W\cos\theta_W}\left(\frac{I_{3,i}}2-Q_i\sin^2\theta_W\right)\text{,}\\
C_{A,i} &=& \frac1{\sin\theta_W\cos\theta_W}\frac{I_{3,i}}2,
\end{eqnarray}
\end{subequations}
where $m_Z$ is the $Z^0$ boson mass, $\theta_W$ the Weinberg angle,
$Q_i$ the fermion charge and $T_{3,i}$ its weak isospin.

The expression for $R_3$ can be obtained from that for $R_2$ with the exchange
 $\mathcal{}A\to-\mathcal{A}$, $\theta_2\to\theta_3$ and $\phi\to\pi-\phi$.

In this case the contribution to $\bar{B}$ is
\begin{equation}
B(\Phi_{B})\mathrm{d}\Phi_{B} {\rm d}x_p \sum_{I\in\left\{{\rm QCDC},{\rm BGF}\right\}}\frac{\alpha_S}{2\pi} S_I,
\end{equation}
where
\begin{subequations}
\begin{eqnarray}
S_{\rm QCDC} &=&  \frac{\frac{x_{B}}{x_{p}} f_{q}(\frac{x_{B}}{x_{p}},Q^{2})}
                      {x_{B} f_{q}(x_{B},Q^{2})}
                 \frac{2+2\ell^2-x_p+3x_p\ell^2+\mathcal{A}\ell\left(1+2x_p\right)}
                      {1+\mathcal{A}\ell+\ell^2}, \\ 
S_{\rm BGF}  &=& -\frac{\frac{x_{B}}{x_{p}} f_{q}(\frac{x_{B}}{x_{p}},Q^{2})}
                      {x_{B} f_{q}(x_{B},Q^{2})}
                 \frac{1+\ell^2+2\left(1-3\ell^2\right)x_p(1-x_p)
                       +2\mathcal{A}\ell\left(1-2x_p(1-x_p)\right)}
                      {1+\mathcal{A}\ell+\ell^2}.\nonumber\\
\end{eqnarray}
\end{subequations}

\subsection{Virtual contribution and collinear remainders}\label{subsec2}

The finite piece of the virtual correction is given by~\cite{Figy:2003nv}
\begin{equation}
\mathrm{d}\sigma_{V}=\frac{C_{F}\alpha_{s}(\mu_{R})}{2\pi}V(x_{B})B(\Phi_{B})\text{,}
\end{equation}
where the finite contribution of the $\mathbf{I}(\epsilon)$ operator of Ref.\,\cite{Catani:1996vz} and the virtual correction is
\begin{equation}
V(x_{B})=-\frac{\pi^{2}}{3}-\frac{9}{2}+\frac{3}{2}\ln\frac{Q^{2}}{\mu_{F}^{2}(1-x_{B})}
+2\ln(1-x_{B})\ln\frac{Q^{2}}{\mu_{F}^{2}}+\ln^{2}(1-x_{B})\text{.}
\end{equation}

The collinear remainders are
\begin{equation}
\mathrm{d}\sigma_{\rm coll}=\frac{C_{F}\alpha_{s}(\mu_{R})}{2\pi}\frac{f^{m}(x_{B},\mu_{F})}{f(x_{B},\mu_{F})}  B(\Phi_{B})\text{,}
\end{equation}
with the modified PDF\footnote{We write the modified PDF for a quark $q$, but a similar expression is valid for an incoming antiquark $\bar{q}$.}
\begin{eqnarray}
f^{m}_{q}(x_{B},\mu_{F})&=&\int_{x_{B}}^{1}\frac{\mathrm{d}x_{p}}{x_{p}}\left\{f_{g}\left(\frac{x_{B}}{x_{p}},\mu_{F}\right)A(x_{p})\right.\notag\\
&+&\left.\left[f_{q}\left(\frac{x_{B}}{x_{p}},\mu_{F}\right)-x_{p}f_{q}(x_{B},\mu_{F})\right]B(x_{p})\right. \notag\\
&+&\left.f_{q}\left(\frac{x_{B}}{x_{p}},\mu_{F}\right)C(x_{p})\right\}\text{,}\ \ \
\end{eqnarray}
where $f_{q}$ and $f_g$ are the quark and gluon PDFs respectively, and
\begin{eqnarray}
A(x_{p})&=&\frac{T_{F}}{C_{F}}\left[x_{p}^{2}+(1-x_{p})^{2}\right]\mathrm{ln}\frac{Q^{2}(1-x_{p})}{\mu_{F}^{2}x_{p}}+2\frac{T_{F}}{C_{F}}x_{p}(1-x_{p})\text{,}\\
B(x_{p})&=&\left[\frac{2}{1-x_{p}}\mathrm{ln}\frac{Q^{2}(1-x_{p})}{\mu_{F}^{2}}-\frac{3}{2}\frac{1}{1-x_{p}}\right]\text{,}\\
C(x_{p})&=&\left[1-x_{p}-\frac{2}{1-x_{p}}\mathrm{ln}x_{p}-(1+x_{p})\mathrm{ln}\frac{Q^{2}(1-x_{p})}{\mu_{F}^{2}x_{p}}\right]\text{.}
\end{eqnarray}
The combined contribution of the finite virtual term and collinear remnants is 
\begin{equation}
\mathrm{d}\sigma_{V+{\rm coll}}=\frac{C_{F}\alpha_{s}(\mu_{R})}{2\pi}\mathcal{V}(\Phi_B)B(\Phi_{B})\text{,}
\end{equation}
where
\begin{equation}
\mathcal{V}(\Phi_{B}) \equiv V(x_{B})+\tilde{V}(x_{B},\mu_{F})\text{,}
\end{equation} 
with $\tilde{V}(x_B,\mu_F)=\frac{f^m(x_B,\mu_F)}{f(x_B,\mu_F)}$.
\subsection{Sampling \boldmath{$\bar{B}$} within \textsf{Herwig++}}\label{subsec3}

Using the results in the previous sections
\begin{eqnarray}
\bar{B}(\Phi_{B})&=&B(\Phi_{B})\left[1+\frac{C_{F}\alpha_{s}(\mu_{R})}{2\pi}\mathcal{V}(\Phi_{B})\right. \notag \\
&+&\left.\sum_{I\in\left\{{\rm QCDC},{\rm BGF}\right\}}\frac{\mathcal{C}_{I}\alpha_{s}(\mu_{R})}{2\pi}\int\left[\mathcal{A}_{I}(\Phi_{B},\Phi_{R})-\mathcal{D}_{I}(\Phi_{B},\Phi_{R})\right]\mathrm{d}\Phi_{R}\right]\text{.} \label{bbar}
\end{eqnarray}
 
For convenience, the radiative variables $\left \{x_{p},z_{p},\phi \right \}$ are transformed into a
new set $\left \{\tilde{x}_{p},z_{p},\tilde{\phi} \right \}$, defined on the interval $[0,1]$, 
such that the radiative phase space is a unit cube. 
The variable $x_{p}$ is redefined as
\begin{equation}
x_{p}=1-\rho^{\frac{1}{1-n}}\text{,}
\end{equation}
where $n$ is fixed, and $\rho$ is the new variable with phase-space limits
\begin{equation}
0<\rho<(1-x_{B})^{1-n}\text{.}
\end{equation}
This change of variable has been made in order to guarantee numerical stability in calculating the integral of $1/(1-x_{p})$.
A further transformation is needed to achieve $\tilde{x}_{p}$: 
\begin{equation}
\rho=(1-x_{B})^{1-n}\tilde{x}_{p}. 
\end{equation}
Finally, the variable $\phi$ is easily redefined as $ \tilde{\phi} = \frac{\phi}{2\pi}$.

The sampling of $\bar{B}(\Phi_{B})$ proceeds in the following way:
\begin{enumerate}
\item generate a leading-order configuration using the standard \textsf{Herwig++} leading-order matrix
element generator, providing the Born variables $\Phi_{B}$ with an associated weight $B(\Phi_{B})$;
\item generate radiative variables, $\Phi_{R}$, by sampling $\bar{B}(\Phi_{B})$,which is parameterized
in terms of the unit cube $(\tilde{x}_{p},z_{p},\tilde{\phi})$, and using the Auto-Compensating 
Divide-and-Conquer~(ACDC) phase-space generator \cite{Lonnblad:2006pt};
\item accepted the leading-order configuration  with a probability proportional to the 
integrand of Eqn.\,\ref{bbar} evaluated at $\left \{\Phi_{B},\Phi_{R}\right \}$. 
\end{enumerate}

In order to treat radiation from both quark lines in the VBF process we randomly select one line which emits the radiation
and generate events in $\Phi\left\{\tilde{x}_{p},z_{p},\tilde{\phi}\right\}$. The symmetry of the process then ensures
that the correct statistical result is obtained by multiplying the correction term in Eqn.\,\ref{bbar} by two.

\section{The generation of the hardest emission}\label{sec3}
The hardest emission is generated using the modified Sudakov form factor, given by the 
product of $\Delta_{R}(p_{T})$ for each channel contributing; this is done replacing
the ratio $R(\Phi_{B},\Phi_{R})/B(\Phi_{B})$ in Eqn.\,\ref{mod_sud} with
\begin{equation}
W_{I}(\Phi_{B},\Phi_{R})=\frac{{R}_{I}(\Phi_{B},\Phi_{R})}{B(\Phi_{B})}\text{.}
\end{equation}
Moreover, we prefer to generate the hardest emission in terms of radiative
variables $\Phi_{R}^{\prime}(x_{\perp},z_{p},\tilde{\phi})$ so that the $\theta$-function
in Eqn.\,\ref{mod_sud} simply gives $x_{\perp}$ as the upper limit of the integral and 
the modified Sudakov form factor, for the channel $I$, becomes
\begin{equation}
\Delta_{R_{I}}(x_{\perp})=\exp \left(-\int_{x_{\perp}}^{x_{\perp}^{\rm{max}}} 
\frac{{\rm d}x^\prime_\perp}{x^{\prime3}_\perp} {\rm d}\tilde{\phi} {\rm d}z_p
\frac{C_{I}\alpha_S}{2\pi} 8z_p(1-z_p)(1-x_p)^2\mathcal{A}_I\right)\text{,}
\end{equation}
where $\frac{Q}2x_{\perp}^{\rm{max}}$ is the maximum value for the transverse momentum. 

The radiative variables are generated using the \textit{veto algorithm}, 
described in \cite{Sjostrand:2006za}. We use the upper bounding function
\begin{equation}
g_{I}=\frac{a_{I}}{x_{\perp}^{3}}\text{,}
\end{equation}
for the integrand which is chosen so that $g_{I}$ can be easily integrated
in $\left\{x_{\perp},x_{\perp}^{\rm{max}}\right\}$.
The generation procedure then proceeds as follows:
\begin{enumerate}
\item $x_{\perp}$ is set to $x_{\perp}^{\rm{max}}$;
\item a new $(\tilde{x}_{p},z_{p},\tilde{\phi})$ is randomly generated according to
\begin{equation}
\Delta^{\rm over}_{R_{I}}(x_{\perp})=\exp \left(-\int_{x_{\perp}}^{x_{\perp}^{\rm{max}}} 
\frac{{\rm d}x^\prime_\perp}{x^{\prime3}_\perp} {\rm d}\tilde{\phi} {\rm d}z_p 
a_{I}\right)\text{,}
\end{equation}
giving\footnote{Here $\mathcal{R}_{i}$ defines a random number in $[0,1]$}
\begin{eqnarray}
x_{\perp}^{2}&=&\frac{1}{\frac{1}{(x_{\perp}^{\rm{max}})^{2}}-\frac{2}{a_{I}}\ln{\mathcal{R}_{1}}}\text{,}\\
z_{p}&=&\mathcal{R}_{2},\\
\tilde{\phi}&=&\mathcal{R}_{3}.
\end{eqnarray}
\item if $\tilde{x}_{p} < 0$ or $\tilde{x}_{p} > 1$, the configuration generated is outside the phase-space boundaries, set $x^{\rm{max}}_\perp$ to $x_\perp$ and return to step 1;
\item if 
\begin{equation}
\frac1{g_{I}}\frac{C_{I}\alpha_S}{2\pi} 8z_p(1-z_p)(1-x_p)^2\mathcal{A}_I > \mathcal{R},
\end{equation}
the configuration is accepted, otherwise set $x^{\rm{max}}_\perp$ to $x_\perp$ and return to step 1.
\end{enumerate}

\section{Truncated and  vetoed parton showers}\label{sec3second}
The \textsf{Herwig++} shower algorithm, \cite{Gieseke:2003rz,Bahr:2008pv}, starts at an
initial scale, given by the colour structure of the hard scattering process, and
evolves down in the evolution variable related to the angular separation of
parton branching products, $\tilde{q}$. The evolution is generated by the 
emission of partons in $1 \to 2$ branching processes and each branching is 
described by a scale, $\tilde{q}$, a light-cone momentum fraction, $z$, and
an azimuthal angle, $\phi$. The latter parameters are used to uniquely define
the momenta of all particles radiated in a shower. However, the \textsf{Herwig++} approach
generally requires some reshuffling of these momenta after the generation
of the parton showers to ensure global energy-momentum conservation.

($N+1$)-body final states associated to the generation of the hardest
emission are first interpreted as a standard \textsf{Herwig++} emission, from 
the $N$-body configuration, specified by the branching 
variables $(\tilde{q}_{h},z_{h},\phi_{h})$. The POWHEG emission is 
performed as a single \textsf{Herwig++} shower as follows:
\begin{enumerate}
\item the truncated shower evolves from the default starting 
scale to $\tilde{q}_{h}$, such that any further emission conserves
the flavour of the emitting parton and has transverse momentum lower
than that of the hardest emission;
\item the hardest emission is forced with shower 
variables $(\tilde{q}_{h},z_{h},\phi_{h})$;
\item the vetoed shower evolves down to the hadronization scale,
vetoing any emission with transverse momentum higher than that of the hardest emission. 
\end{enumerate}   
  
The key feature of this approach is the ability to interpret the hard emission
in terms of the shower variables. In order to do this we first need to
consider the treatment of processes with an initial-state final-state colour
connection, such as DIS or VBF, in the \textsf{Herwig++} parton shower.
In these processes the momenta of the incoming
and outgoing colour connected partons after the parton shower
are first reconstructed from the 
shower variables as described in Ref.\,\cite{Bahr:2008pv}. 
These off-shell momenta are such that energy and momentum is not conserved, so
boosts are applied to the incoming and outgoing momenta 
such that the momentum of the virtual boson is preserved by the showering
process. In DIS and VBF type processes the reconstructed
momenta are boosted to the Breit-frame of the system before the radiation. 
We take $p_b$ to be the momentum of the original incoming parton and $p_c$ to
be the momentum of  the original outgoing parton and $p_a=p_c-p_b$,
therefore in the Breit-frame
\begin{equation}
p_a = Q(1;0,0,-1).
\end{equation}
  We can then construct a set of basis vectors,
\begin{align}
n_1 & = Q(1;0,0,1), & 
n_2 & = Q(1;0,0.-1). 
\end{align}
 
 The momenta of the off-shell incoming parton can be decomposed as
\begin{equation}
q_{\rm in} = \alpha_{\rm in} n_1 + \beta_{\rm in} n_2 +q_{\perp}, 
\end{equation}
 where $\alpha_{\rm in}=\frac{n_2\cdot q_{\rm in}}{n_1\cdot n_2}$, $\beta_{\rm in}=\frac{n_1\cdot q_{\rm in}}{n_1\cdot n_2}$
 and $q_\perp=q_{\rm in}-\alpha_{\rm in} n_1 - \beta_{\rm in} n_2$.
 In order to reconstruct the final-state momentum we first apply a rotation so that
 the momentum of the outgoing jet is
\begin{equation}
q_{\rm out} = \alpha_{\rm out}n_1+\beta_{\rm out}n_2 +q_{\perp},
\end{equation}
where $\beta_{\rm out}=1$ and the
requirement that the virtual mass is preserved gives
$\alpha_{\rm out}=\frac{q^2_{\rm out}+p^2_{\perp}}{2n_1\cdot n_2}$,   
with $q^2_{\perp}=-p^2_{\perp}$.
The momenta of the jets are rescaled such that
\begin{equation}
q'_{\rm in,out} = \alpha_{\rm in,out}k_{\rm in,out}n_1+
\frac{\beta_{\rm in,out}}{k_{\rm in,out}}n_2+q_\perp,
\end{equation}
 which ensures the virtual mass of the partons is preserved.
 The requirement that the momentum of the system is conserved, \textit{i.e.}
\begin{equation}
p_a = q'_{\rm out}-q'_{\rm in} = Q(0,0,-1;0),
\end{equation}
gives
\begin{subequations}
\begin{eqnarray}
\alpha_{\rm in}k_{\rm in}-\alpha_{\rm out}k_{\rm out} &=&\phantom{-} \frac12\text{,}\\
\frac{\beta_{\rm in}}{k_{\rm in}}-\frac{\beta_{\rm out}}{k_{\rm out}} &=&-\frac12.
\end{eqnarray}
\end{subequations}
Once the rescalings have been determined
the jets are transformed using a boost such that 
\begin{equation}
q_{\rm in,out}\stackrel{\mathrm{boost}}{\longrightarrow}q_{\rm in,out}^{\prime}.
\end{equation}

In order to interpret the hard emission in terms of the shower variables we first
calculate the momentum of the off-shell incoming, $q^{\prime {\rm new}}_b$,
or outgoing parton, $q^{\prime {\rm new}}_c$, depending on whether
we are dealing with initial- or final-state radiation. We then compute the 
boost into the Breit-frame of this system and construct the basis vectors $n_{1,2}$
as before which allows us to determine the transverse momentum, $q_\perp$, of
the off-shell incoming parton. In this frame the momenta of the partons
before the shower would be:
\begin{align}
p_b & = \frac{Q}2(1+c;0,0,1+c); & p_c = \frac{Q}2(1+c;0,0,-(1-c)).
\end{align}
The momenta of the off-shell partons before the boost required to conserve
energy and momentum are
\begin{subequations}
\begin{align}
q_b^{\rm new} &= \alpha^{\rm new}_{\rm in}n_1+\beta^{\rm new}_{\rm in}n_2+q_\perp \text{,}&
q_c^{\rm new} &= \alpha^{\rm new}_{\rm out}n_1+\beta^{\rm new}_{\rm out}n_2,
\end{align}
\end{subequations}
where
\begin{subequations}
\begin{align}
\alpha^{\rm new}_{\rm in}  &= \frac{p_b\cdot n_2}{n_1\cdot n_2}, & 
\beta^{\rm new}_{\rm in}   &= \frac{q_b^{\prime2}-q^2_\perp}{2n_1\cdot n_2\alpha^{\rm new}_{\rm in} }, \\
\alpha^{\rm new}_{\rm out} &= \frac{q_c^{\prime2}}{2n_1\cdot n_2\beta^{\rm new}_{\rm out}},  &
\beta^{\rm new}_{\rm out}  &= \frac{p_c\cdot n_1}{n_1\cdot n_2}.
\end{align}
\end{subequations}
The inverse of the boost, which would be applied in the shower to ensure 
energy-momentum conservation, can then be determined and applied to all
the incoming and outgoing partons. These momenta can then be decomposed in terms
of the Sudakov basis used in \textsf{Herwig++}, allowing the
shower variables $(\tilde{q}_h,z_h,\phi_h)$ to be determined.

\section{Results}\label{sec4}
\subsection{Deep Inelastic Scattering}
\begin{figure}
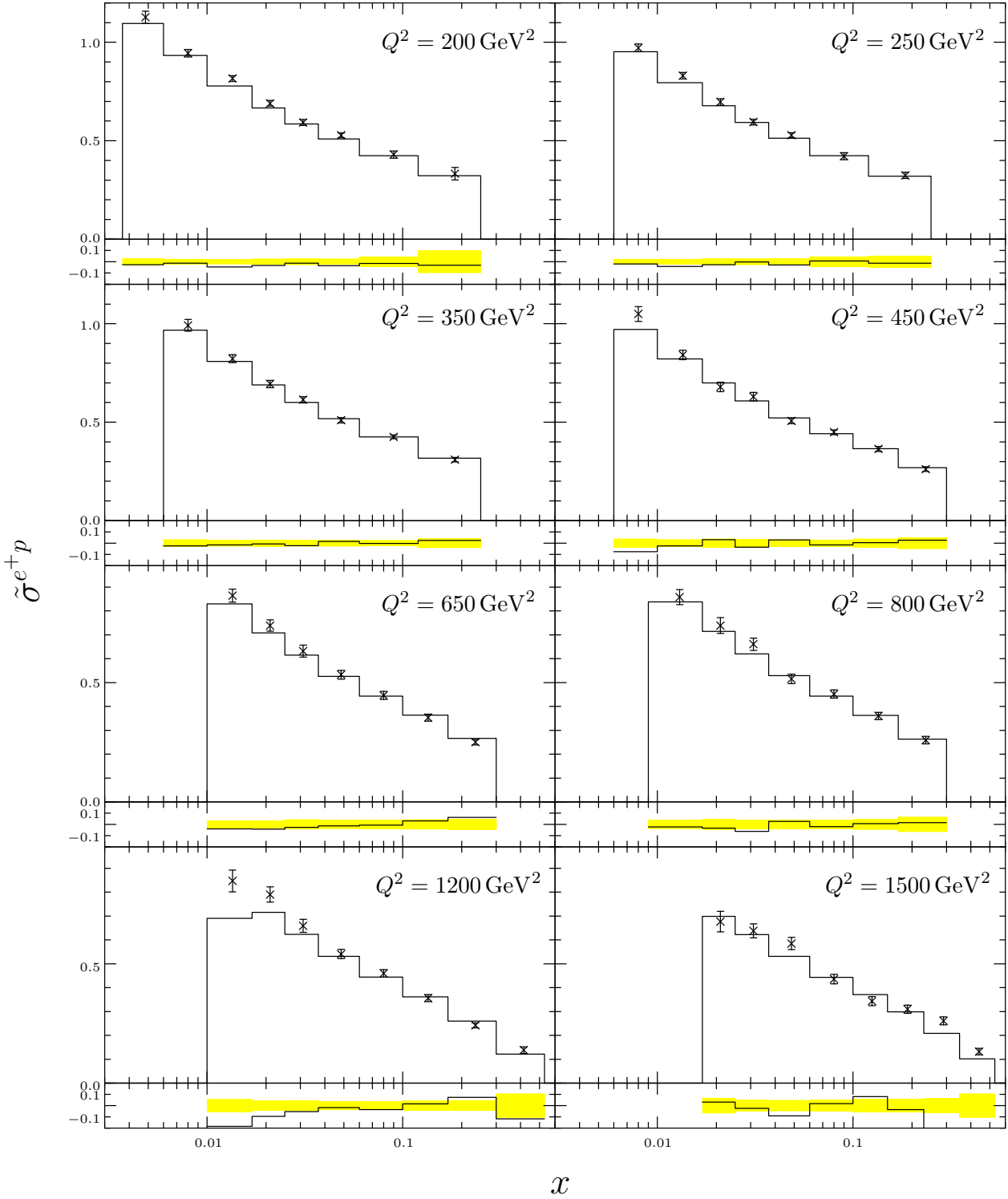

\begin{picture}(0,0)
\Text(168,-20)[c]{\small  $Q^2=200\,{\rm GeV}^2$}
\Text(380,-20)[c]{\small $Q^2=250\,{\rm GeV}^2$}
\Text(168,-150)[c]{\small  $Q^2=350\,{\rm GeV}^2$}
\Text(380,-150)[c]{\small $Q^2=450\,{\rm GeV}^2$}
\Text(168,-285)[c]{\small  $Q^2=650\,{\rm GeV}^2$}
\Text(380,-285)[c]{\small $Q^2=800\,{\rm GeV}^2$}
\Text(168,-418)[c]{\small  $Q^2=1200\,{\rm GeV}^2$}
\Text(380,-418)[c]{\small $Q^2=1500\,{\rm GeV}^2$}
\rText(-40,-270)[c][l]{\Large $\tilde{\sigma}^{e^+p}$}
\rText(210,-560)[c][c]{\Large $x$}
\Text(50,-540)[c]{\tiny $0.01$} 
\Text(142,-540)[c]{\tiny $0.1$} 
\SetOffset(212,0)
\Text(50,-540)[c]{\tiny $0.01$} 
\Text(142,-540)[c]{\tiny $0.1$}

\SetOffset(0,0)
\Text(-1,-513)[r]{\tiny $0.0$}
\Text(-1,-457)[r]{\tiny $0.5$}
\Text(-1,-379)[r]{\tiny $0.0$}
\Text(-1,-323)[r]{\tiny $0.5$}
\Text(-1,-246)[r]{\tiny $0.0$}
\Text(-1,-200)[r]{\tiny $0.5$}
\Text(-1,-154)[r]{\tiny $1.0$}
\Text(-1,-113)[r]{\tiny $0.0$}
\Text(-1,-67)[r]{\tiny $0.5$} 
\Text(-1,-21)[r]{\tiny $1.0$} 
\Text(-1,-118.5)[r]{\tiny $0.1$}
\Text(-1,-130)[r]{\tiny $-0.1$}
\SetOffset(0,-133)
\Text(-1,-118.5)[r]{\tiny $0.1$}
\Text(-1,-130)[r]{\tiny $-0.1$}
\SetOffset(0,-266)
\Text(-1,-118.5)[r]{\tiny $0.1$}
\Text(-1,-130)[r]{\tiny $-0.1$}
\SetOffset(0,-399)
\Text(-1,-118.5)[r]{\tiny $0.1$}
\Text(-1,-130)[r]{\tiny $-0.1$}


\end{picture}

\includegraphics[angle=90,width=\textwidth]{DIS_sigma1_row1.epsi}\\[-1.00mm]
\includegraphics[angle=90,width=\textwidth]{DIS_sigma1_row2.epsi}\\[-1.05mm]
\includegraphics[angle=90,width=\textwidth]{DIS_sigma1_row3.epsi}\\[-1.05mm]
\includegraphics[angle=90,width=\textwidth]{DIS_sigma1_row4.epsi}
\vspace{0.3cm}
\caption{The $e^+p$ reduced cross section $\tilde{\sigma}^{e^+p}$, as a function
         of $x$ at fixed $Q^2$ between $200\,{\rm GeV}^2$ and $1500\,{\rm GeV}^2$.
         The experimental results of Ref.\,\protect\cite{Chekanov:2003yv} are shown as crosses. The
         lower frame shows $({\rm Data}-{\rm Theory})/{\rm Data}$ and the yellow band gives the 
         one sigma variation. The solid~(black) line shows the \textsf{Herwig++} result.}
\label{fig:lowQ2}
\end{figure}

\begin{figure}
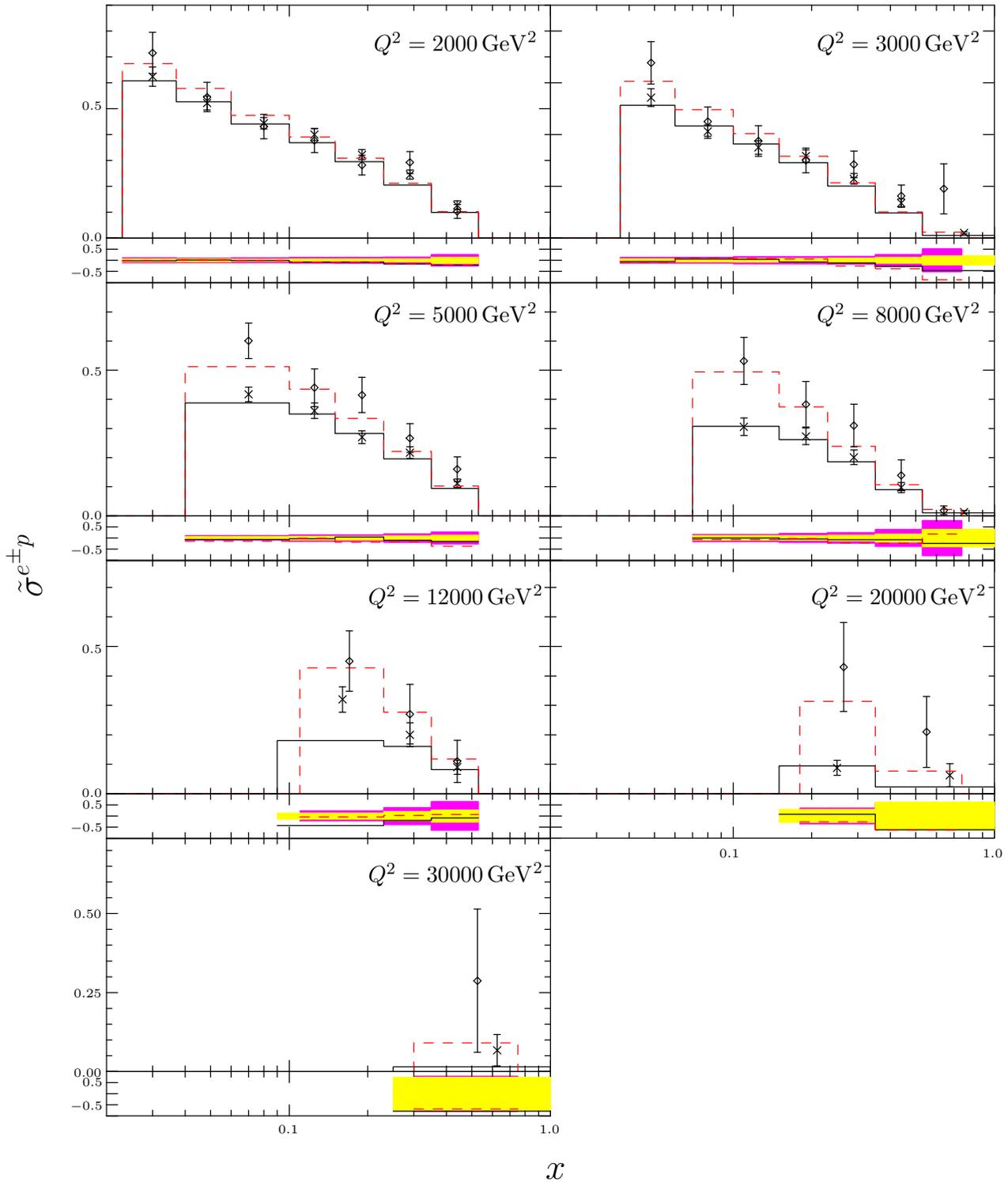

\begin{picture}(0,0)
\Text(168,-20)[c]{\small  $Q^2=2000\,{\rm GeV}^2$}
\Text(380,-20)[c]{\small $Q^2=3000\,{\rm GeV}^2$}
\Text(168,-150)[c]{\small  $Q^2=5000\,{\rm GeV}^2$}
\Text(380,-150)[c]{\small $Q^2=8000\,{\rm GeV}^2$}
\Text(168,-285)[c]{\small  $Q^2=12000\,{\rm GeV}^2$}
\Text(380,-285)[c]{\small $Q^2=20000\,{\rm GeV}^2$}
\Text(168,-418)[c]{\small  $Q^2=30000\,{\rm GeV}^2$}
\rText(-40,-270)[c][l]{\Large $\tilde{\sigma}^{e^\pm p}$}
\rText(210,-560)[c][c]{\Large $x$}
\Text(88,-540)[c]{\tiny $0.1$} 
\Text(213,-540)[c]{\tiny $1.0$}
\SetOffset(212,132)
\Text(87,-540)[c]{\tiny $0.1$} 
\Text(213,-540)[c]{\tiny $1.0$}

\SetOffset(0,0)
\Text(-8,-513)[c]{\tiny $0.00$}
\Text(-8,-475)[c]{\tiny $0.25$}
\Text(-8,-437)[c]{\tiny $0.50$}

\Text(-6,-379)[c]{\tiny $0.0$}
\Text(-6,-310)[c]{\tiny $0.5$}

\Text(-6,-246)[c]{\tiny $0.0$}
\Text(-6,-177)[c]{\tiny $0.5$}

\Text(-6,-113)[c]{\tiny $0.0$}
\Text(-6,-50)[c]{\tiny $0.5$} 
\SetOffset(0,0)
\Text(-1,-118.5)[r]{\tiny $0.5$}
\Text(-1,-130)[r]{\tiny $-0.5$}
\SetOffset(0,-133)
\Text(-1,-118.5)[r]{\tiny $0.5$}
\Text(-1,-130)[r]{\tiny $-0.5$}
\SetOffset(0,-266)
\Text(-1,-118.5)[r]{\tiny $0.5$}
\Text(-1,-130)[r]{\tiny $-0.5$}
\SetOffset(0,-399)
\Text(-1,-118.5)[r]{\tiny $0.5$}
\Text(-1,-130)[r]{\tiny $-0.5$}


\end{picture}

\includegraphics[angle=90,width=\textwidth]{DIS_sigma2_row1.epsi}\\[-0.95mm]
\includegraphics[angle=90,width=\textwidth]{DIS_sigma2_row2.epsi}\\[-0.95mm]
\includegraphics[angle=90,width=\textwidth]{DIS_sigma2_row3.epsi}\\[-0.95mm]
\includegraphics[angle=90,width=0.5023\textwidth]{DIS_sigma2_row4.epsi}
\vspace{0.7cm}
\caption{The $e^\pm p$ reduced cross section $\tilde{\sigma}^{e^\pm p}$, as a function
         of $x$ at fixed $Q^2$ between $200\,{\rm GeV}^2$ and $1500\,{\rm GeV}^2$.
         The experimental results of Ref.\,\protect\cite{Chekanov:2003yv} for $\tilde{\sigma}^{e^+ p}$
         are shown as crosses and the results of Ref.\,\protect\cite{Chekanov:2002ej} 
         for $\tilde{\sigma}^{e^- p}$as diamonds. The
         lower frame shows $({\rm Data}-{\rm Theory})/{\rm Data}$ and the inner~(yellow) band gives the 
         one sigma variation for $\tilde{\sigma}^{e^+ p}$ and the outer~(magenta) band the 
         one sigma variation for $\tilde{\sigma}^{e^- p}$. The solid~(black) line shows the \textsf{Herwig++} result for $\tilde{\sigma}^{e^+ p}$ and
         the dashed~(red) line shower the \textsf{Herwig++} result for $\tilde{\sigma}^{e^- p}$.}
\label{fig:highQ2}
\end{figure}

In order to test our implementation of the POWHEG approach for
deep inelastic scattering we first compared the results from \textsf{Herwig++}
and \textsf{DISENT}~\cite{Catani:1996gg} for the reduced cross section
\begin{equation}
\tilde{\sigma} = \frac{xQ^4}{2\pi\alpha^2Y_+}\frac{{\rm d}^2\sigma}{{\rm d}x{\rm d}Q^2},
\end{equation}
where $y=\frac{Q^2}{xs}$ and $Y_\pm\equiv1\pm(1-y)^2$ and $\alpha$ is the fine-structure
constant. The difference between the \textsf{Herwig++} and \textsf{DISENT} results
divided by the sum of the results is shown as a dashed line in the lower panels
in Figs.\,\ref{fig:lowQ2} and \ref{fig:highQ2} and is always less than one per mille.
In addition Fig.\,\ref{fig:lowQ2} shows the comparison of the \textsf{Herwig++}
result with the results from Ref.\,\cite{Chekanov:2003yv}
and Fig.\,\ref{fig:highQ2} shows the comparison of the \textsf{Herwig++}
result with the results from Refs.\,\cite{Chekanov:2003yv} and \cite{Chekanov:2002ej}.
The excellent agreement with \textsf{DISENT} and the experimental data demonstrates
that the generation of the Born variables and calculation of $\bar{B}$ is correct.
In both cases the PDFs from Ref.\,\cite{Chekanov:2002pv} were used.

\begin{figure}
\begin{picture}(0,0)
\Text(45,10)[c]{\tiny $2.5\,{\rm GeV}^2\!\!<\!Q^2\!\!<\!\!5\,{\rm GeV}^2$}
\Text(130,10)[c]{\tiny $5\,{\rm GeV}^2\!\!<\!Q^2\!\!<\!\!10\,{\rm GeV}^2$}
\Text(215,10)[c]{\tiny $10\,{\rm GeV}^2\!\!<\!Q^2\!\!<\!\!20\,{\rm GeV}^2$}
\Text(300,10)[c]{\tiny $20\,{\rm GeV}^2\!\!<\!Q^2\!\!<\!\!50\,{\rm GeV}^2$}
\Text(385,10)[c]{\tiny $50\,{\rm GeV}^2\!\!<\!Q^2\!\!<\!\!100\,{\rm GeV}^2$}

\Text(45,-439)[c]{\tiny $5\!\times\!10^{-5}\!<\!x\!<\!10^{-4}$}
\Text(45,-332.5)[c]{\tiny $10^{-4}\!<\!x\!<\!2\!\times\!10^{-4}$}
\Text(45,-225)[c]{\tiny $2\!\times\!10^{-4}\!<\!x\!<\!3.5\!\times\!10^{-4}$}
\Text(45,-120)[c]{\tiny $3.5\!\times\!10^{-4}\!<\!x\!<\!10^{-3}$}

\Text(130,-439)[c]{\tiny $10^{-4}\!<\!x\!<\!2\!\times\!10^{-4}$}
\Text(130,-332.5)[c]{\tiny $2\!\times\!10^{-4}\!<\!x\!<\!3.5\!\times\!10^{-4}$}
\Text(130,-225)[c]{\tiny $3.5\!\times\!10^{-4}\!<\!x\!<\!7\!\times\!10^{-4}$}
\Text(130,-120)[c]{\tiny $7\!\times\!10^{-4}\!<\!x\!<\!2\!\times\!10^{-3}$}

\Text(215,-332.5)[c]{\tiny $2\!\times\!10^{-4}\!<\!x\!<\!5\!\times\!10^{-4}$}
\Text(215,-225)[c]{\tiny $5\!\times\!10^{-4}\!<\!x\!<\!8\!\times\!10^{-4}$}
\Text(215,-120)[c]{\tiny $8\!\times\!10^{-4}\!<\!x\!<\!1.5\!\times\!10^{-3}$}
\Text(215,-12 )[c]{\tiny $1.5\!\times\!10^{-3}\!<\!x\!<\!4\!\times\!10^{-2}$}

\Text(300,-225)[c]{\tiny $5\!\times\!10^{-4}\!<\!x\!<\!1.4\!\times\!10^{-3}$}
\Text(300,-120)[c]{\tiny $1.4\!\times\!10^{-3}\!<\!x\!<\!3\!\times\!10^{-3}$}
\Text(300,-12 )[c]{\tiny $3\!\times\!10^{-3}\!<\!x\!<\!\times\!10^{-2}$}

\Text(385,-120)[c]{\tiny $8\!\times\!10^{-4}\!<\!x\!<\!3\!\times\!10^{-3}$}
\Text(385,-12 )[c]{\tiny $3\!\times\!10^{-3}\!<\!x\!<\!2\times\!10^{-2}$} 
\rText(-40,-270)[c][l]{\Large $\frac1N{\rm d}E_T^*/{\rm d}\eta^*/{\rm GeV}$}
\rText(215,-560)[c][c]{\Large $\eta^*$}
\Text(32,-540)[c]{\tiny $0$} 
\Text(77,-540)[c]{\tiny $5$} 
\SetOffset(85,0)
\Text(32,-540)[c]{\tiny $0$} 
\Text(77,-540)[c]{\tiny $5$} 
\SetOffset(170,0)
\Text(32,-435)[c]{\tiny $0$} 
\Text(77,-435)[c]{\tiny $5$} 
\SetOffset(255,0)
\Text(32,-328)[c]{\tiny $0$} 
\Text(77,-328)[c]{\tiny $5$} 
\SetOffset(340,0)
\Text(32,-221)[c]{\tiny $0$} 
\Text(77,-221)[c]{\tiny $5$} 

\SetOffset(0,0)
\Text(-5,-513)[c]{\tiny $0$}
\Text(-5,-469)[c]{\tiny $2$}
\Text(-5,-407)[c]{\tiny $0$}
\Text(-5,-363)[c]{\tiny $2$}
\Text(-5,-300)[c]{\tiny $0$}
\Text(-5,-257)[c]{\tiny $2$}
\Text(-5,-194)[c]{\tiny $0$}
\Text(-5,-151)[c]{\tiny $2$} 

\SetOffset(0,-399)
\Text(-1,-118.5)[r]{\tiny $0.2$}
\Text(-1,-130)[r]{\tiny $-0.2$}
\SetOffset(0,-293)
\Text(-1,-118.5)[r]{\tiny $0.2$}
\Text(-1,-130)[r]{\tiny $-0.2$}
\SetOffset(0,-187)
\Text(-1,-118.5)[r]{\tiny $0.2$}
\Text(-1,-130)[r]{\tiny $-0.2$}
\SetOffset(0,-80)
\Text(-1,-118.5)[r]{\tiny $0.2$}
\Text(-1,-130)[r]{\tiny $-0.2$}

\SetOffset(170,0)
\Text(-5,-87)[c]{\tiny $0$}
\Text(-5,-44)[c]{\tiny $2$}
\SetOffset(170,27)
\Text(-1,-118.5)[r]{\tiny $0.2$}
\Text(-1,-130)[r]{\tiny $-0.2$}
\SetOffset(0,0)
\Text(325,-395)[l]{Hw}
\Text(325,-418)[l]{Hw++}
\Text(325,-441)[l]{Hw++, ME correction}
\Text(325,-464)[l]{Hw++, POWHEG}
\end{picture}

\hspace{5.98cm}\includegraphics[angle=90,width=0.6012\textwidth]{DIS_ET1_row1.epsi}\\[-0.85mm]
\includegraphics[angle=90,width=\textwidth]{DIS_ET1_row2.epsi}\\[-0.85mm]
\includegraphics[angle=90,width=0.8007\textwidth]{DIS_ET1_row3.epsi}\\[-0.85mm]
\includegraphics[angle=90,width=0.6014\textwidth]{DIS_ET1_row4.epsi}\\[-0.85mm]
\includegraphics[angle=90,width=0.4021\textwidth]{DIS_ET1_row5.epsi}

\vspace{-5cm}
\hspace{9.2cm}\includegraphics[angle=90,width=0.15\textwidth]{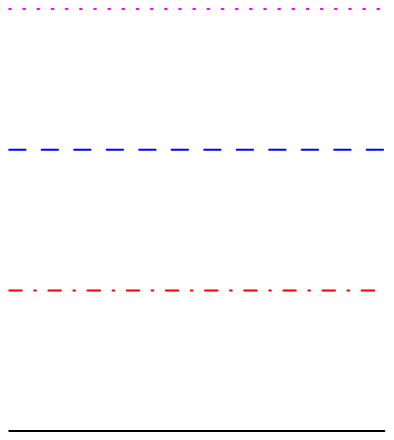}\\[-0.55mm]
\vspace{3cm}
\caption{The inclusive transverse energy flow $\frac1N{\rm d}E_T^*/{\rm d}\eta^*$ at different values of
         $x$ and $Q^2$ for the low $Q^2$ sample from \protect\cite{Adloff:1999ws}. The
         lower frame shows $({\rm Data}-{\rm Theory})/{\rm Data}$ and the yellow band gives the 
         one sigma variation.}
\label{fig:lowQ2ET}
\end{figure}

\begin{figure}
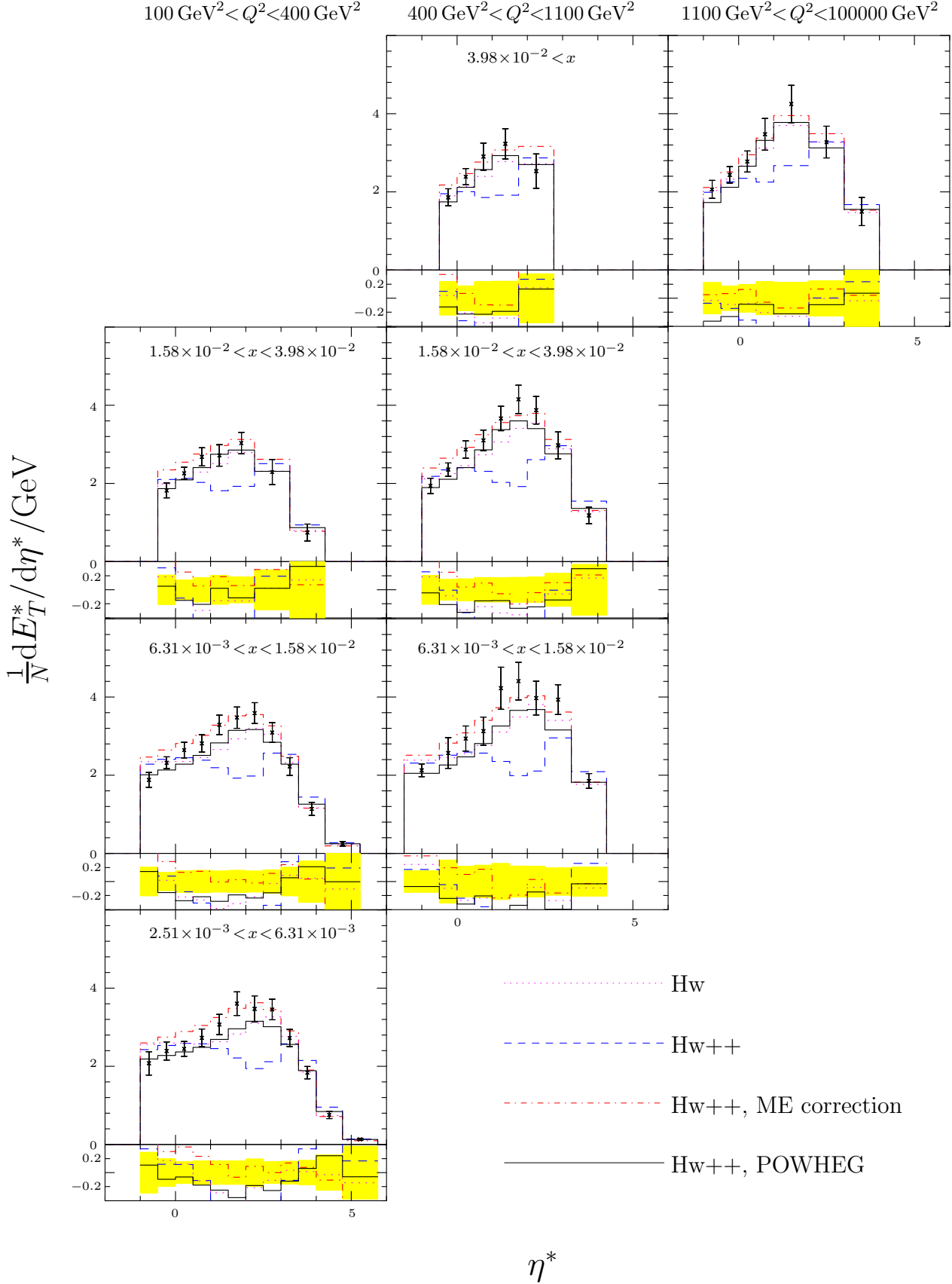

\begin{picture}(0,0)
\Text(75,10)[c]{\footnotesize $100\,{\rm GeV}^2\!\!<\!Q^2\!\!<\!\!400\,{\rm GeV}^2$}
\Text(210,10)[c]{\footnotesize $400\,{\rm GeV}^2\!\!<\!Q^2\!\!<\!\!1100\,{\rm GeV}^2$}
\Text(355,10)[c]{\footnotesize $1100\,{\rm GeV}^2\!\!<\!Q^2\!\!<\!\!100000\,{\rm GeV}^2$}

\Text(75,-452)[c]{\scriptsize $2.51\!\times\!10^{-3}\!<\!x\!<\!6.31\!\times\!10^{-3}$}
\Text(75,-309)[c]{\scriptsize $6.31\!\times\!10^{-3}\!<\!x\!<\!1.58\!\times\!10^{-2}$}
\Text(75,-160)[c]{\scriptsize $1.58\!\times\!10^{-2}\!<\!x\!<\!3.98\!\times\!10^{-2}$}

\Text(210,-309)[c]{\scriptsize $6.31\!\times\!10^{-3}\!<\!x\!<\!1.58\!\times\!10^{-2}$}
\Text(210,-160)[c]{\scriptsize $1.58\!\times\!10^{-2}\!<\!x\!<\!3.98\!\times\!10^{-2}$}
\Text(210,-12)[c]{\scriptsize $3.98\!\times\!10^{-2}\!<\!x$}

\rText(-40,-270)[c][l]{\Large $\frac1N{\rm d}E_T^*/{\rm d}\eta^*/{\rm GeV}$}
\rText(215,-620)[c][c]{\Large $\eta^*$}

\Text(36,-593)[c]{\tiny $0$} 
\Text(125,-593)[c]{\tiny $5$} 
\SetOffset(142,146)
\Text(36,-593)[c]{\tiny $0$} 
\Text(125,-593)[c]{\tiny $5$} 
\SetOffset(284,438)
\Text(36,-593)[c]{\tiny $0$} 
\Text(125,-593)[c]{\tiny $5$} 

\SetOffset(0,0)

\Text(-5,-559)[c]{\tiny $0$}
\Text(-5,-518)[c]{\tiny $2$}
\Text(-5,-480)[c]{\tiny $4$}
\Text(-1,-565.5)[r]{\tiny $0.2$}
\Text(-1,-580)[r]{\tiny $-0.2$}
\SetOffset(0,146)
\Text(-5,-559)[c]{\tiny $0$}
\Text(-5,-518)[c]{\tiny $2$}
\Text(-5,-480)[c]{\tiny $4$}
\Text(-1,-565.5)[r]{\tiny $0.2$}
\Text(-1,-580)[r]{\tiny $-0.2$}
\SetOffset(0,292)
\Text(-5,-559)[c]{\tiny $0$}
\Text(-5,-518)[c]{\tiny $2$}
\Text(-5,-480)[c]{\tiny $4$}
\Text(-1,-565.5)[r]{\tiny $0.2$}
\Text(-1,-580)[r]{\tiny $-0.2$}
\SetOffset(142,439)
\Text(-5,-559)[c]{\tiny $0$}
\Text(-5,-518)[c]{\tiny $2$}
\Text(-5,-480)[c]{\tiny $4$}
\Text(-1,-565.5)[r]{\tiny $0.2$}
\Text(-1,-580)[r]{\tiny $-0.2$}
\SetOffset(0,0)
\Text(285,-478)[l]{Hw}
\Text(285,-509)[l]{Hw++}
\Text(285,-540)[l]{Hw++, ME correction}
\Text(285,-570)[l]{Hw++, POWHEG}
\end{picture}

\hspace{4.97cm}\includegraphics[angle=90,width=0.6667\textwidth]{DIS_ET2_row1.epsi}\\[-0.70mm]
\includegraphics[angle=90,width=0.666666\textwidth]{DIS_ET2_row2.epsi}\\[-0.70mm]
\includegraphics[angle=90,width=0.666666\textwidth]{DIS_ET2_row3.epsi}\\[-0.75mm]
\includegraphics[angle=90,width=0.3353\textwidth]{DIS_ET2_row4.epsi}

\vspace{-4cm}
\hspace{7cm}\includegraphics[angle=90,width=0.20\textwidth]{lines}\\[-0.55mm]
\vspace{1.2cm}
\caption{The inclusive transverse energy flow $\frac1N{\rm d}E_T^*/{\rm d}\eta^*$ at different values of
         $x$ and $Q^2$ for the high $Q^2$ sample from \protect\cite{Adloff:1999ws}. The
         lower frame shows $({\rm Data}-{\rm Theory})/{\rm Data}$ and the yellow band gives the 
         one sigma variation.}
\label{fig:highQ2ET}
\end{figure}

In order to study the real emission we compare the results of \textsf{Herwig++}
with the measurements of the transverse energy flow in DIS from
Ref.~\cite{Adloff:1999ws} which are sensitive to the treatment of 
hard radiation in angular-ordered parton showers \cite{Seymour:1994ti}. The
comparison of \textsf{Herwig++} with the low and high $Q^2$ samples from
Ref.~\cite{Adloff:1999ws} are shown in Figs.\,\ref{fig:lowQ2ET} and
\ref{fig:highQ2ET}, respectively. In addition to the \textsf{Herwig++}
result, with and without the POWHEG correction, we have included the result
of the \textsf{FORTRAN} \textsf{HERWIG}\cite{Corcella:2002jc,Corcella:2000bw}
and \textsf{Herwig++} with a matrix element correction based on the 
approach of Ref.\,\cite{Seymour:1994ti}. These results
clearly show that without a correction to describe hard QCD radiation there is
a deficit of emissions between $1<\eta^*<3$ which is remedied by using either
the POWHEG approach or a traditional matrix element correction. In general 
the POWHEG approach gives slightly less radiation than the matrix element
due to the Sudakov suppression of radiation which is neglected in the matrix
element correction approach and is in the best agreement with the
experimental results. In these plots we have tuned the mass parameter
for the splitting of soft beam remnant clusters in \textsf{Herwig++} to 0.5\,GeV from the
\textsf{HERWIG} value of 1\,GeV. The transverse energy flow in DIS is most sensitive to 
this parameter and the original \textsf{HERWIG} value was tuned to older transverse energy
flow data.

\subsection{Higgs Boson Production via Vector Boson Fusion}
The typical feature of the VBF process at hadron colliders is the presence of two
 forward tagging jets. At leading order, they correspond to the two scattered
 quarks in the hard process and their observation, together with the properties
 of the Higgs boson decay products, is vital for the suppression of
 backgrounds \cite{Rainwater:1998kj, Plehn:1999xi, Rainwater:1999sd, Kauer:2000hi, Rainwater:1997dg, Eboli:2000ze, Cavalli:2002vs}.
 The tagging jet distributions must be known precisely to gain a good estimate
 of the Higgs boson couplings: comparison of the Higgs boson production rate
 with the tagging jet cross section, within cuts, determines the Higgs boson
 couplings, \cite{Zeppenfeld:2000td, Belyaev:2002ua}, and the uncertainties 
 of the measured couplings are determined by the theoretical error of the
 cross section. At next-to-leading-order, the tagging jet distributions
 are sufficient to estimate size and uncertainties of the higher order
 QCD corrections, because the Higgs boson does not induce spin
 correlations in the phase space of its decay products.

\begin{figure}
\centering
\includegraphics[angle=90, width=0.9\textwidth]{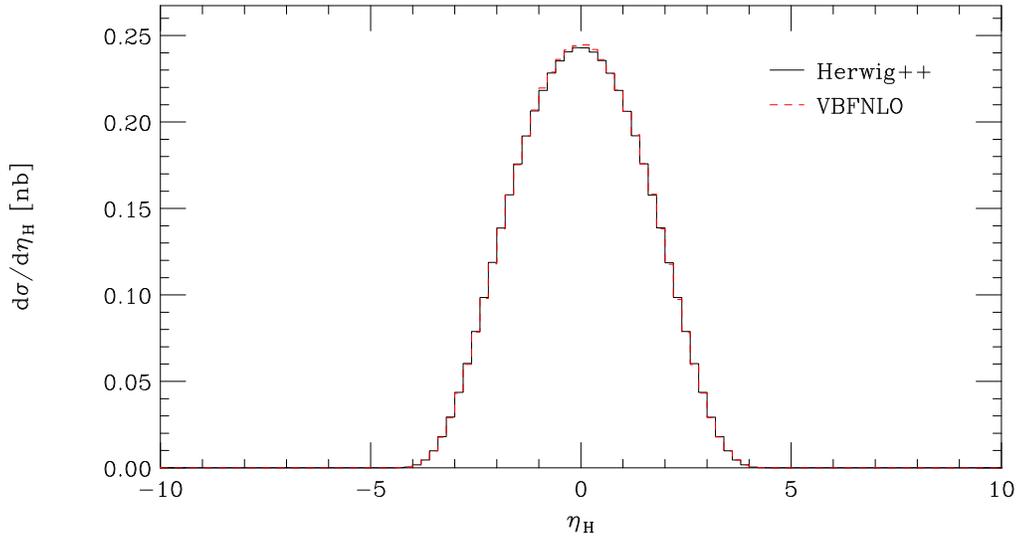}
\caption{Rapidity of the Higgs boson at NLO. Results from \textsf{Herwig++} (solid black line) are compared to the one from \textsf{VBFNLO} (dashed red line).}\label{rap_h}
\end{figure}

A detailed analysis of jet distributions has been realized in the present work and the results are shown in this section. A preliminary step has been the validation of the $\bar{B}$ function, by comparing the NLO differential cross section as function of the rapidity of the stable Higgs boson given by \textsf{Herwig++} and \textsf{VBFNLO}, as shown in Fig.~\ref{rap_h}.  

However, at next-to-leading-order we can either encounter two jets, with one of them composed of two partons (recombination effects), or three jets corresponding to well-separated partons. As for LHC data, an algorithm to select two tagging jets is needed. There are two possibilities \cite{Figy:2003nv}:
\begin{enumerate}
\item \textit{$p_{T}$-method}: the two tagging jets are the two highest $p_{T}$ jets in the event;
\item \textit{$E$-method}: the two tagging jets are the two highest energy jets in the event.
\end{enumerate}

We follow the $p_{T}$-method and jets are defined according to the $k_{T}$ algorithm by using the \textsf{FastJet}
package \cite{Cacciari:2005hq}. Cuts need to be chosen to reduce the effect of backgrounds and we follow the ones introduced in \cite{Figy:2003nv}. Tagging jets are required to have transverse momentum, $p_{T}$, and rapidity, $y_{j}$, fulfilling the following cuts:
\begin{equation}
p_{T}\ge 20~{\rm GeV}\text{,}\hspace{1 cm} | y_{j}|\le 4.5\text{.} 
\end{equation}  
Moreover we generate the Higgs boson decay in $\tau^{+}\tau^{-}$ isotropically and require that the produced leptons have transverse momentum, $p_{T_{\tau^{(+,-)}}}$, and pseudorapidity, $\eta_{\tau^{(+,-)}}$, so that
\begin{equation}
p_{T_{\tau^{(+,-)}}}\ge 20~\rm{GeV}\text{,}\hspace{1 cm} | \eta_{\tau^{(+,-)}}|\le 2.5\text{.}
\end{equation}
In addition, we require that jet-lepton separation in the rapidity-azimuthal angle plane satisfies 
\begin{equation}
\Delta R_{j\tau^{(+,-)}}\ge 0.6\text{,}
\end{equation}
and the tau leptons to lie between the two tagging jets in rapidity
\begin{equation}
y_{j,\rm{min}}<\eta_{\tau^{(+,-)}}<y_{j,{\rm{min}}}\text{.}
\end{equation}
Backgrounds to VBF are significantly suppressed if the two tagging jets are well separated in rapidity; therefore, we require
\begin{equation}
| y_{j_{1}}-y_{j_{2}}|>4\text{.}\label{rapsepcut}
\end{equation}
The factorization and the renormalization scale are chosen to be equal to the mass of the Higgs boson, $m_{H}=120~\rm{GeV}$.  The other relevant electroweak parameters are
\begin{equation}
M_{W}=80.3980\rm{ GeV}\text{,}\hspace{0.5 cm}M_{Z}=91.1876\rm{ GeV}\text{,}\hspace{0.5 cm}\alpha_{em}=0.007556\text{,}\hspace{0.5 cm}\rm{sin}^{2}\theta_{W}=0.222646\text{,}
\end{equation}
and the weak coupling is computed as $g=\sqrt{4\alpha_{em}/\rm{sin}\theta_{W}}$. The parton distribution functions are chosen to be the \rm{CTEQ6M} set \cite{Pumplin:2002vw,Tung:2002vr}.

The analysis provides the comparison of distributions for \textsf{POWHEG} implementation (solid black curve) and LO simulation (dashed red curve) of \textsf{Herwig++} parton shower together with \textsf{VBFNLO} NLO differential cross section (dotted blue curve).

\begin{figure}
\includegraphics[angle=90,width=0.5\textwidth]{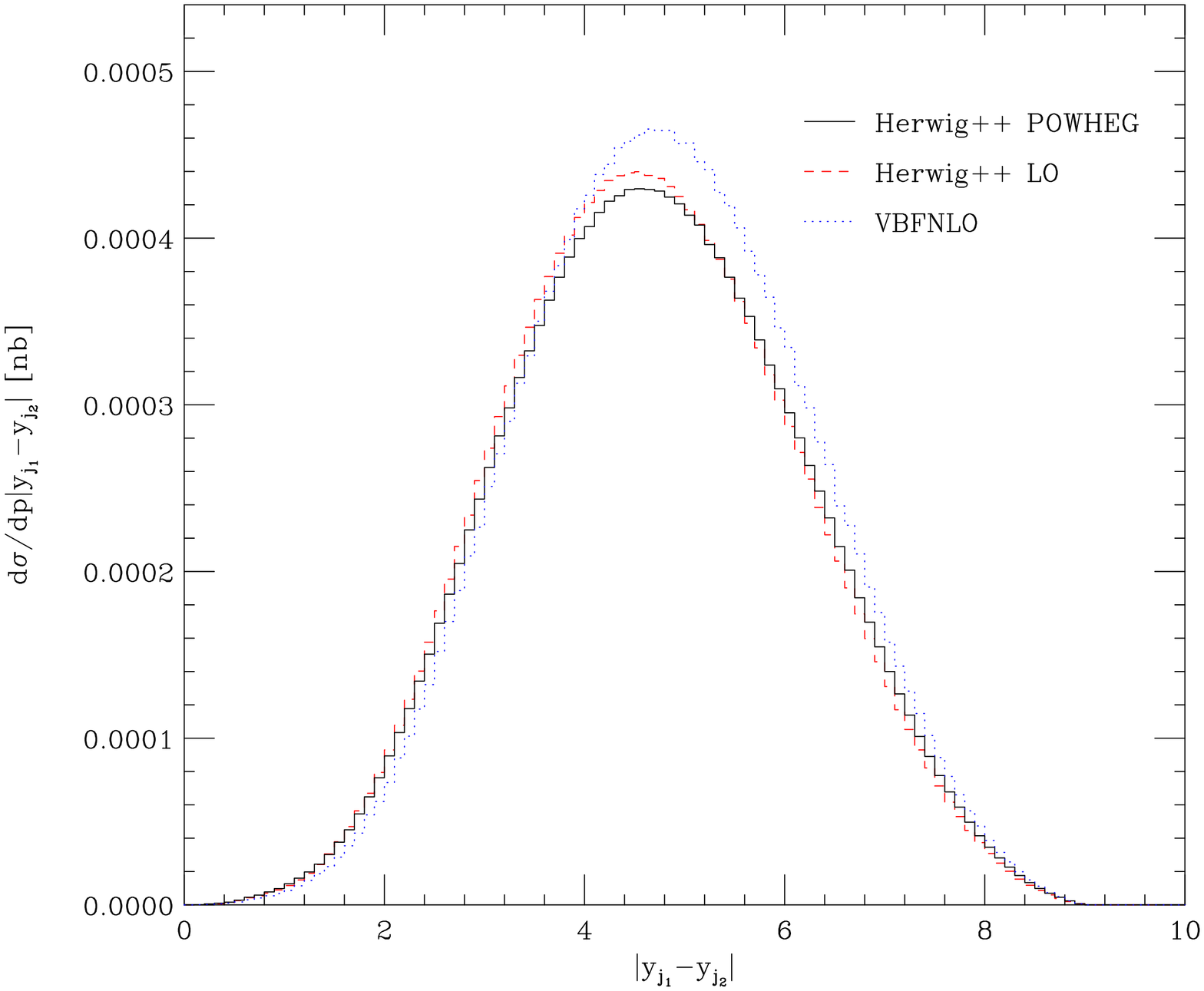} 
\includegraphics[angle=90,width=0.5\textwidth]{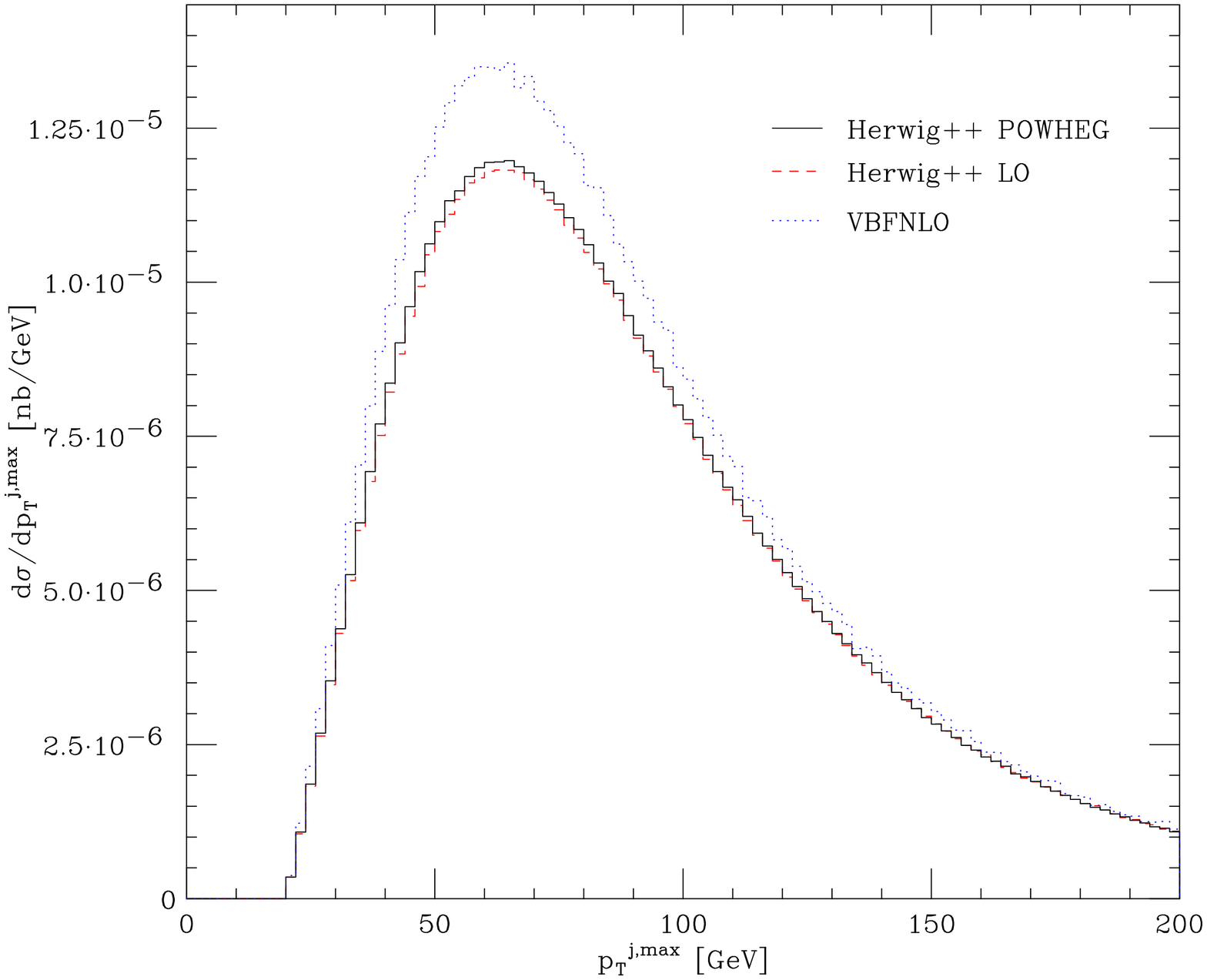}\\[-0.55mm]
\vspace{0.3cm}
\caption{Difference of rapidity (left panel) and higher $p_{T}$ (right panel) distribution of the two tagging jets. In the left panel we have excluded  the cut in Eqn.~\ref{rapsepcut}.}
\label{VBF_fig1}
\end{figure}

In Fig.~\ref{VBF_fig1} we present the differential cross section as a function of the rapidity separation and higher $p_{T}$ of the two tagging jets.
 In the left panel we have excluded the cut in Eqn.~\ref{rapsepcut}. The cross sections show a peak at a rapidity
 of around $5$ in the left panel and a transverse momentum of $70$ GeV
in the right panel. The \textsf{POWHEG} implementation leaves the cross section with respect to the $p_T$ of the hardest jet unchanged,
while it modifies the difference of rapidity distribution respect to the LO simulation of the \textsf{Herwig++} parton shower: the peak is slightly lower and shifted to a higher value of the rapidity difference.   

\begin{figure}
\includegraphics[angle=90,width=0.5\textwidth]{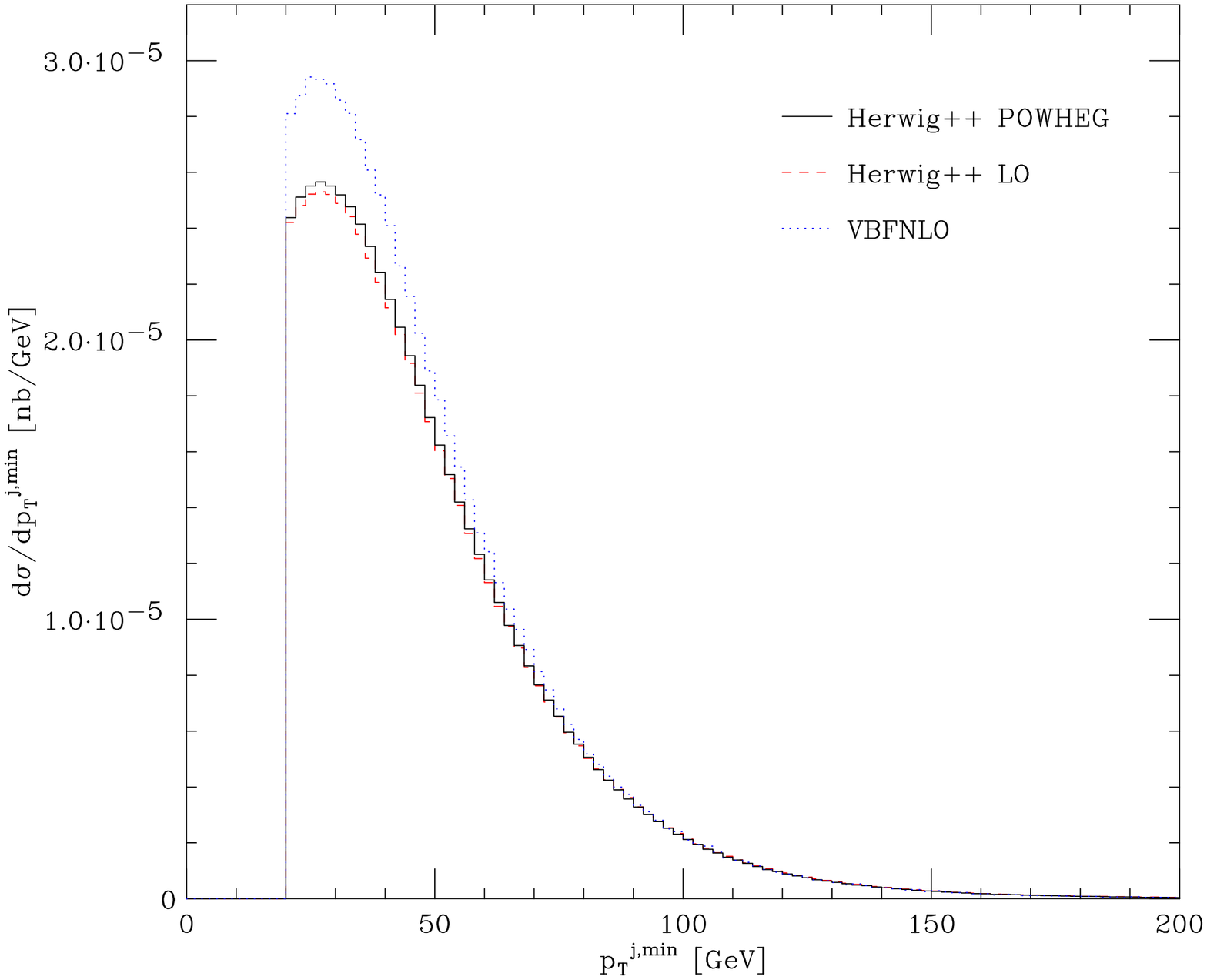}
\includegraphics[angle=90,width=0.5\textwidth]{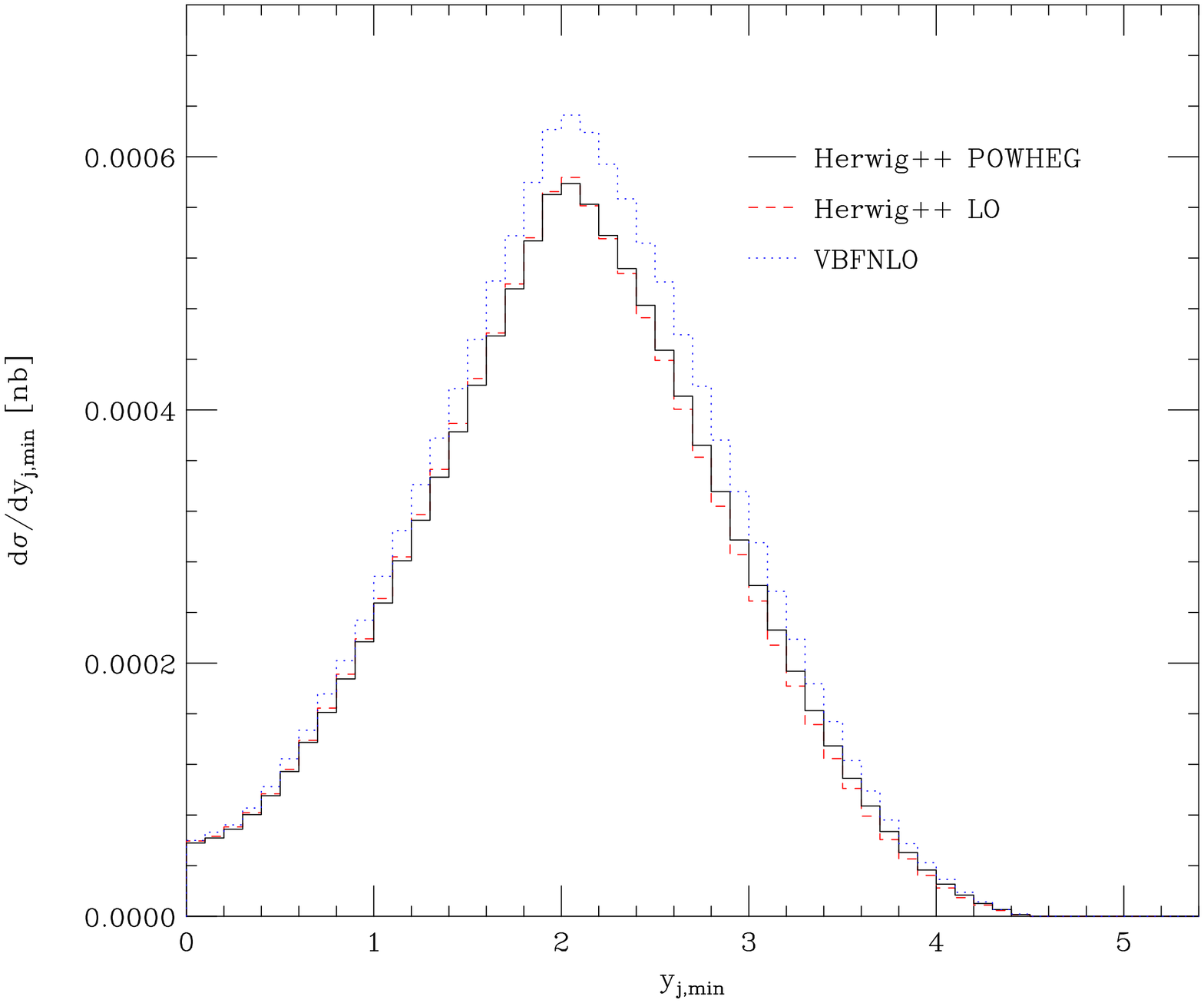}\\[-0.55mm]
\vspace{0.3cm}
\caption{Distributions of smaller transverse momentum, $p_{T}^{j,{\rm min}}$, (left panel) and smaller rapidity, $y_{j,{\rm min}}$, (right panel) of the two tagging jets.}
\label{VBF_fig2}
\end{figure}

In Fig.~\ref{VBF_fig2} we plot the cross section with respect to the transverse momentum (left panel) and
 rapidity (right panel) of the softer of the two tagging jets. The transverse momentum distribution shows a peak
 around $30$ GeV and the rapidity around $2$. The \textsf{Herwig++} shower provides a similar description at LO and NLO accuracy.
 
\begin{figure}
\includegraphics[angle=90,width=0.5\textwidth]{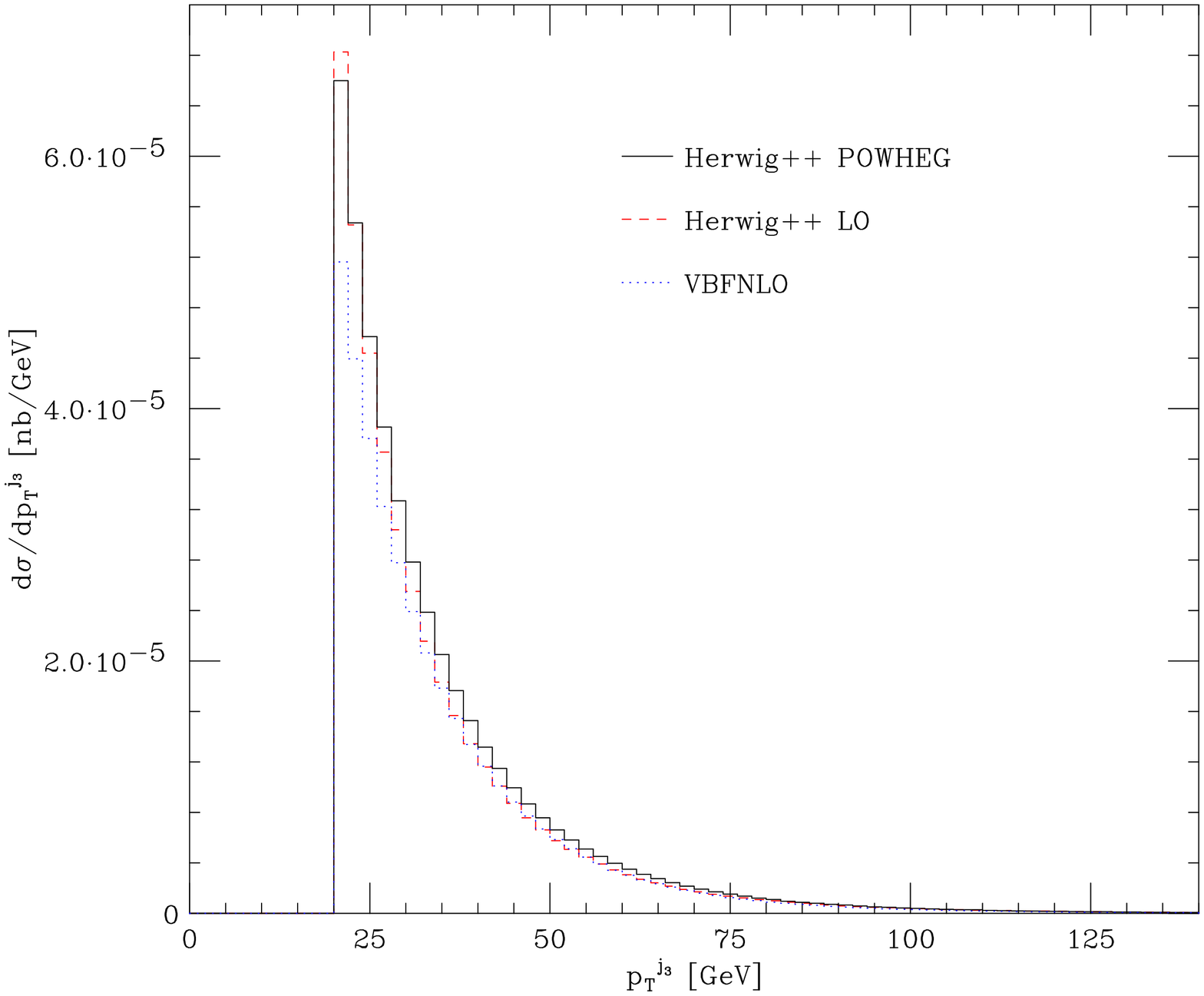}
\includegraphics[angle=90,width=0.5\textwidth]{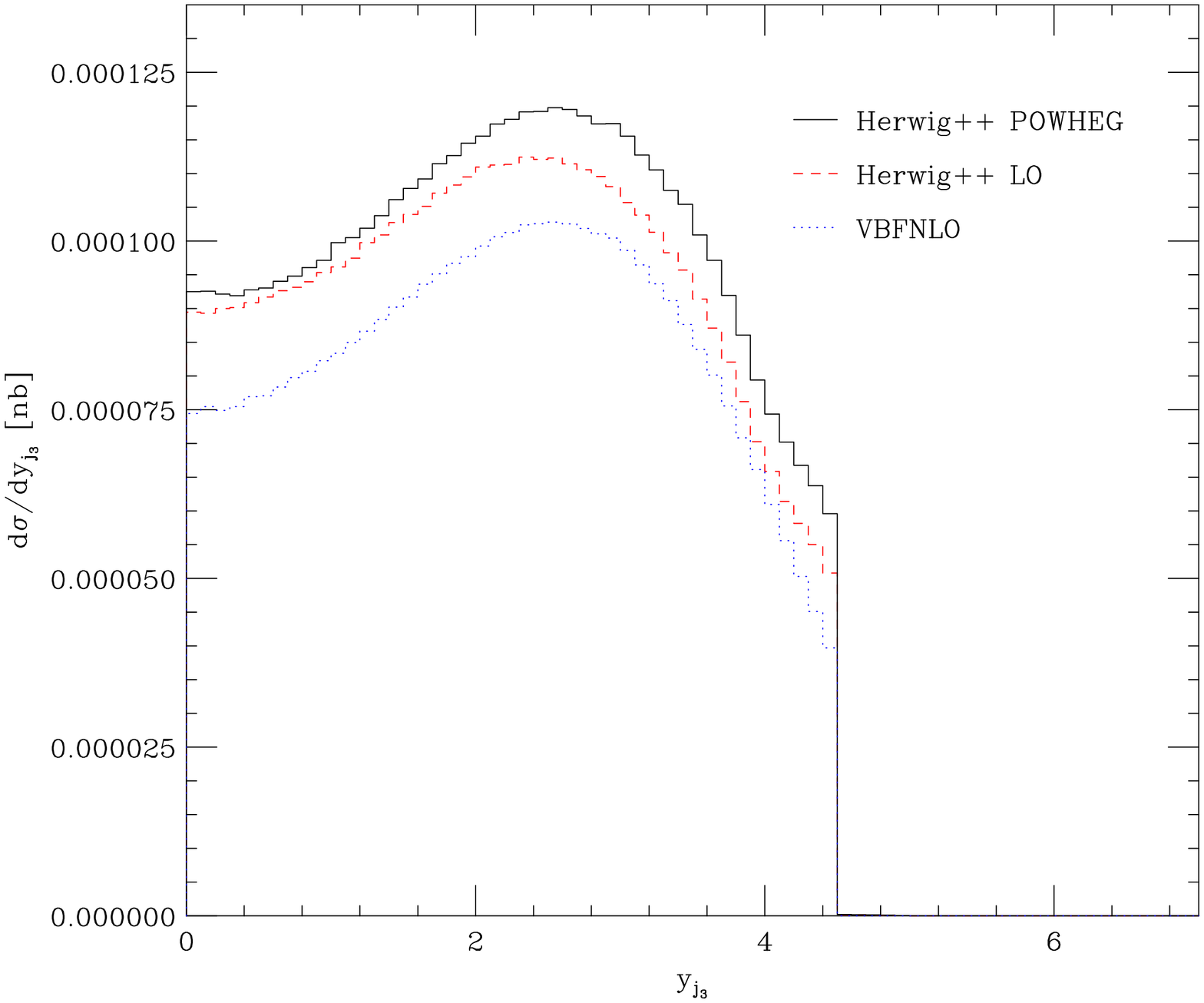}\\[-0.55mm]
\vspace{0.3cm}
\caption{Transverse momentum distribution of the third jet ($p_{T}^{j_{3}}$) is plotted in the left panel and the rapidity distribution of the same jet ($y_{j_3}$) in the right one.}
\label{VBF_fig3} 
\end{figure}

The transverse momentum and rapidity distributions of the third jet are plotted
in Fig.~\ref{VBF_fig3} in the left and right panel respectively. As would be expected here we see
a harder spectrum for the third jet in the POWHEG approach which is now simulated using
the real emission matrix element rather than the shower approximation.

In these plots we see that the \textsf{Herwig++} results lie below the fixed NLO results for
distributions involving the two leading jets as a result of the subsequent parton shower, unlike the results of
Ref.\,\cite{Nason:2009ai} where there is little difference between the POWHEG and fixed order results. This
difference exists at both leading order and in the POWHEG approach in \textsf{Herwig++} and is
a result of the different shower algorithm and kinematic reconstruction in \textsf{Herwig++}. Given the
excellent description of the related DIS data, compared with the previous \textsf{HERWIG} shower algorithm, 
this is an important difference in the two approaches and worthy of further study.

\section{Conclusion}\label{sec5}

In this work the \textsf{POWHEG} NLO matching scheme has been implemented in
 the \textsf{Herwig++} Monte Carlo event generator for DIS and Higgs boson production
 via vector boson fusion. For both hard subprocesses the $\bar{B}$ function has been
 calculated following the general approach provided in \cite{Seymour:1994df} which
enables other colour-singlet VBF production processes to be easily included.
 The simulation contains a full treatment of the truncated shower, which is
 needed for the production of wide angle, soft radiation in angular-ordered parton showers.

For the DIS implementation we find the cross sections to be in good agreement with the experimental results of Refs.\cite{Chekanov:2003yv, Chekanov:2002ej, Adloff:1999ws}. Our results show that the \textsf{POWHEG} approach correctly populates the so-called \textit{dead zone}, as it appears in the transverse energy flow distributions at high $Q^2$.

For the VBF Higgs boson production we have shown different jet distributions after imposing typical cuts which are required to remove the effects of backgrounds. We find that the \textsf{POWHEG} implementation does improve the rapidity separation distribution for the two tagging jets and of $p_{T}$ and rapidity for the third hardest jet, while it mainly leaves the other distributions of the two tagging jets unchanged respect to the LO simulations within the \textsf{Herwig++} parton shower.
  
The lack of data prevents us from comparing the jet distributions with experimental results. However, the present work together with the Higgs production via gluon fusion and the Higgs-strahlung simulations, which were already implemented in \textsf{Herwig++~2.3} \cite{Hamilton:2009za}, will provide an important tool for analyzing the upcoming results at the LHC.   

\bibliography{Herwig++}

\providecommand{\href}[2]{#2}\begingroup\raggedright\begin{thebibliography}{100}

\bibitem{Englert:1964et}
F.~Englert and R.~Brout, {\it {B}roken {S}ymmetry and the mass of {G}auge
  {V}ector {M}esons},  {\em Phys. Rev. Lett.} {\bf 13} (1964) 321--322.

\bibitem{Higgs:1964pj}
P.~W. Higgs, {\it {B}roken {S}ymmetries and the {M}asses of {G}auge {B}osons},
  {\em Phys. Rev. Lett.} {\bf 13} (1964) 508--509.

\bibitem{Guralnik:1964eu}
G.~S. Guralnik, C.~R. Hagen, and T.~W.~B. Kibble, {\it {G}lobal {C}onservation
  {L}aws and {M}assless {P}articles},  {\em Phys. Rev. Lett.} {\bf 13} (1964)
  585--587.

\bibitem{Kibble:1967sv}
T.~W.~B. Kibble, {\it {Symmetry breaking in non-Abelian gauge theories}},  {\em
  Phys. Rev.} {\bf 155} (1967) 1554--1561.

\bibitem{Zeppenfeld:2000td}
D.~Zeppenfeld, R.~Kinnunen, A.~Nikitenko, and E.~Richter-Was, {\it {Measuring
  Higgs boson couplings at the LHC}},  {\em Phys. Rev.} {\bf D62} (2000)
  013009, [\href{http://xxx.lanl.gov/abs/hep-ph/0002036}{{\tt
  hep-ph/0002036}}].

\bibitem{Belyaev:2002ua}
A.~Belyaev and L.~Reina, {\it {$pp\to t\bar t H$, $H\to\tau^+\tau^-$: toward a
  model independent determination of the Higgs boson couplings at the LHC}},
  {\em JHEP} {\bf 08} (2002) 041,
  [\href{http://xxx.lanl.gov/abs/hep-ph/0205270}{{\tt hep-ph/0205270}}].

\bibitem{Georgi:1977gs}
H.~M. Georgi, S.~L. Glashow, M.~E. Machacek, and D.~V. Nanopoulos, {\it {Higgs
  Bosons from Two Gluon Annihilation in Proton Proton Collisions}},  {\em Phys.
  Rev. Lett.} {\bf 40} (1978) 692.

\bibitem{Rainwater:1998kj}
D.~L. Rainwater, D.~Zeppenfeld, and K.~Hagiwara, {\it {Searching for
  $H\to\tau^+\tau^-$ in weak boson fusion at the LHC}},  {\em Phys. Rev.} {\bf
  D59} (1999) 014037, [\href{http://xxx.lanl.gov/abs/hep-ph/9808468}{{\tt
  hep-ph/9808468}}].

\bibitem{Plehn:1999xi}
T.~Plehn, D.~L. Rainwater, and D.~Zeppenfeld, {\it {A method for identifying
  $H\to\tau^+\tau^-\to e^\pm\mu^\mp$+ missing $p_T$ at the CERN LHC}},  {\em
  Phys. Rev.} {\bf D61} (2000) 093005,
  [\href{http://xxx.lanl.gov/abs/hep-ph/9911385}{{\tt hep-ph/9911385}}].

\bibitem{Rainwater:1999sd}
D.~L. Rainwater and D.~Zeppenfeld, {\it {Observing $H \to W^{(*)}W^{(*)} \to
  e^\pm \mu^\mp /\!\!\!{p}_T$ in weak boson fusion with dual forward jet
  tagging at the CERN LHC}},  {\em Phys. Rev.} {\bf D60} (1999) 113004,
  [\href{http://xxx.lanl.gov/abs/hep-ph/9906218}{{\tt hep-ph/9906218}}].

\bibitem{Kauer:2000hi}
N.~Kauer, T.~Plehn, D.~L. Rainwater, and D.~Zeppenfeld, {\it {$H\to WW$ as the
  discovery mode for a light Higgs boson}},  {\em Phys. Lett.} {\bf B503}
  (2001) 113--120, [\href{http://xxx.lanl.gov/abs/hep-ph/0012351}{{\tt
  hep-ph/0012351}}].

\bibitem{Rainwater:1997dg}
D.~L. Rainwater and D.~Zeppenfeld, {\it {Searching for $H\to\gamma\gamma$ in
  weak boson fusion at the LHC}},  {\em JHEP} {\bf 12} (1997) 005,
  [\href{http://xxx.lanl.gov/abs/hep-ph/9712271}{{\tt hep-ph/9712271}}].

\bibitem{Eboli:2000ze}
O.~J.~P. Eboli and D.~Zeppenfeld, {\it {Observing an invisible Higgs boson}},
  {\em Phys. Lett.} {\bf B495} (2000) 147--154,
  [\href{http://xxx.lanl.gov/abs/hep-ph/0009158}{{\tt hep-ph/0009158}}].

\bibitem{Cavalli:2002vs}
D.~Cavalli {\em et.~al.}, {\it {The Higgs working group: Summary report}},
  \href{http://xxx.lanl.gov/abs/hep-ph/0203056}{{\tt hep-ph/0203056}}.

\bibitem{Han:1991ia}
T.~Han and S.~Willenbrock, {\it {QCD correction to the p p $\to$ W H and Z H
  total cross- sections}},  {\em Phys. Lett.} {\bf B273} (1991) 167--172.

\bibitem{Djouadi:1991tka}
A.~Djouadi, M.~Spira, and P.~M. Zerwas, {\it {Production of Higgs bosons in
  proton colliders: QCD corrections}},  {\em Phys. Lett.} {\bf B264} (1991)
  440--446.

\bibitem{Spira:1995rr}
M.~Spira, A.~Djouadi, D.~Graudenz, and P.~M. Zerwas, {\it {Higgs boson
  production at the LHC}},  {\em Nucl. Phys.} {\bf B453} (1995) 17--82,
  [\href{http://xxx.lanl.gov/abs/hep-ph/9504378}{{\tt hep-ph/9504378}}].

\bibitem{Dawson:1990zj}
S.~Dawson, {\it {Radiative corrections to Higgs boson production}},  {\em Nucl.
  Phys.} {\bf B359} (1991) 283--300.

\bibitem{Giele:2002hx}
W.~Giele {\em et.~al.}, {\it {The QCD / SM working group: Summary report}},
  \href{http://xxx.lanl.gov/abs/hep-ph/0204316}{{\tt hep-ph/0204316}}.

\bibitem{Catani:2001ic}
S.~Catani, D.~de~Florian, and M.~Grazzini, {\it {Higgs production in hadron
  collisions: Soft and virtual QCD corrections at NNLO}},  {\em JHEP} {\bf 05}
  (2001) 025, [\href{http://xxx.lanl.gov/abs/hep-ph/0102227}{{\tt
  hep-ph/0102227}}].

\bibitem{Harlander:2001is}
R.~V. Harlander and W.~B. Kilgore, {\it {Soft and virtual corrections to $pp\to
  H+X$ at NNLO}},  {\em Phys. Rev.} {\bf D64} (2001) 013015,
  [\href{http://xxx.lanl.gov/abs/hep-ph/0102241}{{\tt hep-ph/0102241}}].

\bibitem{Harlander:2002wh}
R.~V. Harlander and W.~B. Kilgore, {\it {Next-to-next-to-leading order Higgs
  production at hadron colliders}},  {\em Phys. Rev. Lett.} {\bf 88} (2002)
  201801, [\href{http://xxx.lanl.gov/abs/hep-ph/0201206}{{\tt
  hep-ph/0201206}}].

\bibitem{Anastasiou:2002yz}
C.~Anastasiou and K.~Melnikov, {\it {Higgs boson production at hadron colliders
  in NNLO QCD}},  {\em Nucl. Phys.} {\bf B646} (2002) 220--256,
  [\href{http://xxx.lanl.gov/abs/hep-ph/0207004}{{\tt hep-ph/0207004}}].

\bibitem{Ravindran:2003um}
V.~Ravindran, J.~Smith, and W.~L. van Neerven, {\it {NNLO corrections to the
  total cross section for Higgs boson production in hadron hadron collisions}},
   {\em Nucl. Phys.} {\bf B665} (2003) 325--366,
  [\href{http://xxx.lanl.gov/abs/hep-ph/0302135}{{\tt hep-ph/0302135}}].

\bibitem{Buckley:2011ms}
A.~Buckley {\em et.~al.}, {\it {General-purpose event generators for LHC
  physics}},  \href{http://xxx.lanl.gov/abs/1101.2599}{{\tt arXiv:1101.2599}}.

\bibitem{Sjostrand:1986hx}
T.~Sjostrand and M.~Bengtsson, {\it {The Lund Monte Carlo for Jet Fragmentation
  and e+ e- Physics. Jetset Version 6.3: An Update}},  {\em Comput. Phys.
  Commun.} {\bf 43} (1987) 367.

\bibitem{Bengtsson:1987rw}
M.~Bengtsson and T.~Sjostrand, {\it {P}arton {S}howers in {L}eptoproduction
  {E}vents},  {\em Z. Phys.} {\bf C37} (1988) 465.

\bibitem{Norrbin:2000uu}
E.~Norrbin and T.~Sjostrand, {\it {QCD radiation off heavy particles}},  {\em
  Nucl. Phys.} {\bf B603} (2001) 297--342,
  [\href{http://xxx.lanl.gov/abs/hep-ph/0010012}{{\tt hep-ph/0010012}}].

\bibitem{Miu:1998ju}
G.~Miu and T.~Sjostrand, {\it {$W$ production in an improved parton shower
  approach}},  {\em Phys. Lett.} {\bf B449} (1999) 313--320,
  [\href{http://xxx.lanl.gov/abs/hep-ph/9812455}{{\tt hep-ph/9812455}}].

\bibitem{Corcella:2000bw}
G.~Corcella {\em et.~al.}, {\it {HERWIG} 6: An event generator for {H}adron
  {E}mission {R}eactions with {I}nterfering {G}luons (including supersymmetric
  processes)},  {\em JHEP} {\bf 01} (2001) 010,
  [\href{http://xxx.lanl.gov/abs/hep-ph/0011363}{{\tt hep-ph/0011363}}].

\bibitem{Corcella:2002jc}
G.~Corcella {\em et.~al.}, {\it {HERWIG} 6.5 {R}elease {N}ote},
  \href{http://xxx.lanl.gov/abs/hep-ph/0210213}{{\tt hep-ph/0210213}}.

\bibitem{Seymour:1991xa}
M.~H. Seymour, {\it {Photon radiation in final state parton showering}},  {\em
  Z. Phys.} {\bf C56} (1992) 161--170.

\bibitem{Seymour:1994ti}
M.~H. Seymour, {\it {Matrix element corrections to parton shower simulation of
  deep inelastic scattering}}, . Contributed to 27th International Conference
  on High Energy Physics (ICHEP), Glasgow, Scotland, 20-27 Jul 1994.

\bibitem{Corcella:1998rs}
G.~Corcella and M.~H. Seymour, {\it {Matrix element corrections to parton
  shower simulations of heavy quark decay}},  {\em Phys. Lett.} {\bf B442}
  (1998) 417--426, [\href{http://xxx.lanl.gov/abs/hep-ph/9809451}{{\tt
  hep-ph/9809451}}].

\bibitem{Corcella:1999gs}
G.~Corcella and M.~H. Seymour, {\it {Initial state radiation in simulations of
  vector boson production at hadron colliders}},  {\em Nucl. Phys.} {\bf B565}
  (2000) 227--244, [\href{http://xxx.lanl.gov/abs/hep-ph/9908388}{{\tt
  hep-ph/9908388}}].

\bibitem{Seymour:1994df}
M.~H. Seymour, {\it {M}atrix {E}lement {C}orrections to {P}arton {S}hower
  {A}lgorithms},  {\em Comp. Phys. Commun.} {\bf 90} (1995) 95--101,
  [\href{http://xxx.lanl.gov/abs/hep-ph/9410414}{{\tt hep-ph/9410414}}].

\bibitem{Seymour:1994we}
M.~H. Seymour, {\it {A Simple prescription for first order corrections to quark
  scattering and annihilation processes}},  {\em Nucl. Phys.} {\bf B436} (1995)
  443--460, [\href{http://xxx.lanl.gov/abs/hep-ph/9410244}{{\tt
  hep-ph/9410244}}].

\bibitem{Gieseke:2003hm}
S.~Gieseke, A.~Ribon, M.~H. Seymour, P.~Stephens, and B.~Webber, {\it
  {Herwig++} 1.0: {A}n {E}vent {G}enerator for ${\rm e}^+{\rm e}^-$
  {A}nnihilation},  {\em JHEP} {\bf 02} (2004) 005,
  [\href{http://xxx.lanl.gov/abs/hep-ph/0311208}{{\tt hep-ph/0311208}}].

\bibitem{Gieseke:2004af}
S.~Gieseke, {\it {The new Monte Carlo event generator Herwig++}},
  \href{http://xxx.lanl.gov/abs/hep-ph/0408034}{{\tt hep-ph/0408034}}.

\bibitem{Hamilton:2006ms}
K.~Hamilton and P.~Richardson, {\it {A} {S}imulation of {QCD} {R}adiation in
  {T}op {Q}uark {D}ecays},  {\em JHEP} {\bf 02} (2007) 069,
  [\href{http://xxx.lanl.gov/abs/hep-ph/0612236}{{\tt hep-ph/0612236}}].

\bibitem{Gieseke:2006ga}
S.~Gieseke {\em et.~al.}, {\it Herwig++ 2.0 {R}elease {N}ote},
  \href{http://xxx.lanl.gov/abs/hep-ph/0609306}{{\tt hep-ph/0609306}}.

\bibitem{Bahr:2008tx}
M.~Bahr {\em et.~al.}, {\it {Herwig++ 2.2 Release Note}},
  \href{http://xxx.lanl.gov/abs/0804.3053}{{\tt arXiv:0804.3053}}.

\bibitem{Gieseke:2011na}
S.~Gieseke, D.~Grellscheid, K.~Hamilton, A.~Papaefstathiou, S.~Platzer, {\em
  et.~al.}, {\it {Herwig++ 2.5 Release Note}},
  \href{http://xxx.lanl.gov/abs/1102.1672}{{\tt arXiv:1102.1672}}.

\bibitem{Catani:2001cc}
S.~Catani, F.~Krauss, R.~Kuhn, and B.~R. Webber, {\it {QCD} {M}atrix {E}lements
  + {P}arton {S}howers},  {\em JHEP} {\bf 11} (2001) 063,
  [\href{http://xxx.lanl.gov/abs/hep-ph/0109231}{{\tt hep-ph/0109231}}].

\bibitem{Krauss:2002up}
F.~Krauss, {\it {Matrix elements and parton showers in hadronic interactions}},
   {\em JHEP} {\bf 08} (2002) 015,
  [\href{http://xxx.lanl.gov/abs/hep-ph/0205283}{{\tt hep-ph/0205283}}].

\bibitem{Lonnblad:2001iq}
L.~Lonnblad, {\it {Correcting the colour-dipole cascade model with fixed order
  matrix elements}},  {\em JHEP} {\bf 05} (2002) 046,
  [\href{http://xxx.lanl.gov/abs/hep-ph/0112284}{{\tt hep-ph/0112284}}].

\bibitem{Schalicke:2005nv}
A.~Schalicke and F.~Krauss, {\it {Implementing the ME+PS merging algorithm}},
  {\em JHEP} {\bf 07} (2005) 018,
  [\href{http://xxx.lanl.gov/abs/hep-ph/0503281}{{\tt hep-ph/0503281}}].

\bibitem{Krauss:2005re}
F.~Krauss, A.~Schalicke, and G.~Soff, {\it {APACIC++ 2.0: A Parton cascade in
  C++}},  {\em Comput. Phys. Commun.} {\bf 174} (2006) 876--902,
  [\href{http://xxx.lanl.gov/abs/hep-ph/0503087}{{\tt hep-ph/0503087}}].

\bibitem{Lavesson:2005xu}
N.~Lavesson and L.~Lonnblad, {\it {W + jets matrix elements and the dipole
  cascade}},  {\em JHEP} {\bf 07} (2005) 054,
  [\href{http://xxx.lanl.gov/abs/hep-ph/0503293}{{\tt hep-ph/0503293}}].

\bibitem{Mrenna:2003if}
S.~Mrenna and P.~Richardson, {\it {Matching matrix elements and parton showers
  with HERWIG and PYTHIA}},  {\em JHEP} {\bf 05} (2004) 040,
  [\href{http://xxx.lanl.gov/abs/hep-ph/0312274}{{\tt hep-ph/0312274}}].

\bibitem{Mangano:2002ea}
M.~L. Mangano, M.~Moretti, F.~Piccinini, R.~Pittau, and A.~D. Polosa, {\it
  {ALPGEN, a generator for hard multiparton processes in hadronic collisions}},
   {\em JHEP} {\bf 07} (2003) 001,
  [\href{http://xxx.lanl.gov/abs/hep-ph/0206293}{{\tt hep-ph/0206293}}].

\bibitem{Alwall:2007fs}
J.~Alwall {\em et.~al.}, {\it {Comparative study of various algorithms for the
  merging of parton showers and matrix elements in hadronic collisions}},  {\em
  Eur. Phys. J.} {\bf C53} (2008) 473--500,
  [\href{http://xxx.lanl.gov/abs/0706.2569}{{\tt arXiv:0706.2569}}].

\bibitem{Hoeche:2009rj}
S.~Hoeche, F.~Krauss, S.~Schumann, and F.~Siegert, {\it {QCD matrix elements
  and truncated showers}},  {\em JHEP} {\bf 05} (2009) 053,
  [\href{http://xxx.lanl.gov/abs/0903.1219}{{\tt arXiv:0903.1219}}].

\bibitem{Hamilton:2009ne}
K.~Hamilton, P.~Richardson, and J.~Tully, {\it {A modified CKKW matrix element
  merging approach to angular-ordered parton showers}},  {\em JHEP} {\bf 11}
  (2009) 038, [\href{http://xxx.lanl.gov/abs/0905.3072}{{\tt
  arXiv:0905.3072}}].

\bibitem{Frixione:2002ik}
S.~Frixione and B.~R. Webber, {\it {M}atching {NLO} {QCD} {C}omputations and
  {P}arton {S}hower {S}imulations},  {\em JHEP} {\bf 06} (2002) 029,
  [\href{http://xxx.lanl.gov/abs/hep-ph/0204244}{{\tt hep-ph/0204244}}].

\bibitem{Frixione:2010wd}
S.~Frixione, F.~Stoeckli, P.~Torrielli, B.~R. Webber, and C.~D. White, {\it
  {The MC@NLO 4.0 Event Generator}},
  \href{http://xxx.lanl.gov/abs/1010.0819}{{\tt arXiv:1010.0819}}.

\bibitem{Frixione:2005vw}
S.~Frixione, E.~Laenen, P.~Motylinski, and B.~R. Webber, {\it {Single-top
  Production in MC@NLO}},  {\em JHEP} {\bf 03} (2006) 092,
  [\href{http://xxx.lanl.gov/abs/hep-ph/0512250}{{\tt hep-ph/0512250}}].

\bibitem{Frixione:2007zp}
S.~Frixione, E.~Laenen, P.~Motylinski, and B.~R. Webber, {\it {Angular
  Correlations of Lepton Pairs from Vector Boson and Top Quark Decays in Monte
  Carlo Simulations}},  {\em JHEP} {\bf 04} (2007) 081,
  [\href{http://xxx.lanl.gov/abs/hep-ph/0702198}{{\tt hep-ph/0702198}}].

\bibitem{Frixione:2008yi}
S.~Frixione, E.~Laenen, P.~Motylinski, B.~R. Webber, and C.~D. White, {\it
  {Single-top hadroproduction in association with a W boson}},  {\em JHEP} {\bf
  07} (2008) 029, [\href{http://xxx.lanl.gov/abs/0805.3067}{{\tt
  arXiv:0805.3067}}].

\bibitem{LatundeDada:2007jg}
O.~Latunde-Dada, {\it {Herwig++ Monte Carlo At Next-To-Leading Order for e+e-
  annihilation and lepton pair production}},  {\em JHEP} {\bf 11} (2007) 040,
  [\href{http://xxx.lanl.gov/abs/0708.4390}{{\tt arXiv:0708.4390}}].

\bibitem{LatundeDada:2009rr}
O.~Latunde-Dada, {\it {MC@NLO for the hadronic decay of Higgs bosons in
  associated production with vector bosons}},  {\em JHEP} {\bf 05} (2009) 112,
  [\href{http://xxx.lanl.gov/abs/0903.4135}{{\tt arXiv:0903.4135}}].

\bibitem{Papaefstathiou:2009sr}
A.~Papaefstathiou and O.~Latunde-Dada, {\it {NLO production of $W'$ bosons at
  hadron colliders using the MC@NLO and POWHEG methods}},  {\em JHEP} {\bf 07}
  (2009) 044, [\href{http://xxx.lanl.gov/abs/0901.3685}{{\tt
  arXiv:0901.3685}}].

\bibitem{Torrielli:2010aw}
P.~Torrielli and S.~Frixione, {\it {Matching NLO QCD computations with PYTHIA
  using MC@NLO}},  {\em JHEP} {\bf 1004} (2010) 110,
  [\href{http://xxx.lanl.gov/abs/1002.4293}{{\tt arXiv:1002.4293}}].

\bibitem{Frixione:2010ra}
S.~Frixione, F.~Stoeckli, P.~Torrielli, and B.~R. Webber, {\it {NLO QCD
  corrections in Herwig++ with MC@NLO}},  {\em JHEP} {\bf 1101} (2011) 053,
  [\href{http://xxx.lanl.gov/abs/1010.0568}{{\tt arXiv:1010.0568}}].

\bibitem{Nason:2004rx}
P.~Nason, {\it A new method for combining {NLO} {QCD} with shower {M}onte
  {C}arlo algorithms},  {\em JHEP} {\bf 11} (2004) 040,
  [\href{http://xxx.lanl.gov/abs/hep-ph/0409146}{{\tt hep-ph/0409146}}].

\bibitem{Frixione:2007vw}
S.~Frixione, P.~Nason, and C.~Oleari, {\it {M}atching {NLO} {QCD} computations
  with {P}arton {S}hower simulations: the {POWHEG} method},  {\em JHEP} {\bf
  11} (2007) 070, [\href{http://xxx.lanl.gov/abs/0709.2092}{{\tt 0709.2092}}].

\bibitem{Nason:2006hfa}
P.~Nason and G.~Ridolfi, {\it {A Positive-Weight Next-to-leading-Order Monte
  Carlo for Z pair Hadroproduction}},  {\em JHEP} {\bf 08} (2006) 077,
  [\href{http://xxx.lanl.gov/abs/hep-ph/0606275}{{\tt hep-ph/0606275}}].

\bibitem{Frixione:2007nw}
S.~Frixione, P.~Nason, and G.~Ridolfi, {\it {A Positive-Weight
  Next-to-Leading-Order Monte Carlo for Heavy Flavour Hadroproduction}},  {\em
  JHEP} {\bf 09} (2007) 126, [\href{http://xxx.lanl.gov/abs/0707.3088}{{\tt
  arXiv:0707.3088}}].

\bibitem{LatundeDada:2006gx}
O.~Latunde-Dada, S.~Gieseke, and B.~Webber, {\it {A} {P}ositive-{W}eight
  {N}ext-to-{L}eading-{O}rder {M}onte {C}arlo for $e^+e-$ annihilation to
  hadrons},  {\em JHEP} {\bf 02} (2007) 051,
  [\href{http://xxx.lanl.gov/abs/hep-ph/0612281}{{\tt hep-ph/0612281}}].

\bibitem{Alioli:2008gx}
S.~Alioli, P.~Nason, C.~Oleari, and E.~Re, {\it {NLO vector-boson production
  matched with shower in POWHEG}},  {\em JHEP} {\bf 07} (2008) 060,
  [\href{http://xxx.lanl.gov/abs/0805.4802}{{\tt arXiv:0805.4802}}].

\bibitem{Hamilton:2008pd}
K.~Hamilton, P.~Richardson, and J.~Tully, {\it {A Positive-Weight
  Next-to-Leading Order Monte Carlo Simulation of Drell-Yan Vector Boson
  Production}},  \href{http://xxx.lanl.gov/abs/0806.0290}{{\tt
  arXiv:0806.0290}}.

\bibitem{Alioli:2008tz}
S.~Alioli, P.~Nason, C.~Oleari, and E.~Re, {\it {NLO Higgs boson production via
  gluon fusion matched with shower in POWHEG}},  {\em JHEP} {\bf 04} (2009)
  002, [\href{http://xxx.lanl.gov/abs/0812.0578}{{\tt arXiv:0812.0578}}].

\bibitem{Hamilton:2009za}
K.~Hamilton, P.~Richardson, and J.~Tully, {\it {A Positive-Weight
  Next-to-Leading Order Monte Carlo Simulation for Higgs Boson Production}},
  {\em JHEP} {\bf 04} (2009) 116,
  [\href{http://xxx.lanl.gov/abs/0903.4345}{{\tt arXiv:0903.4345}}].

\bibitem{Alioli:2009je}
S.~Alioli, P.~Nason, C.~Oleari, and E.~Re, {\it {NLO single-top production
  matched with shower in POWHEG: s- and t-channel contributions}},  {\em JHEP}
  {\bf 09} (2009) 111, [\href{http://xxx.lanl.gov/abs/0907.4076}{{\tt
  arXiv:0907.4076}}].

\bibitem{Nason:2009ai}
P.~Nason and C.~Oleari, {\it {NLO Higgs boson production via vector-boson
  fusion matched with shower in POWHEG}},
  \href{http://xxx.lanl.gov/abs/0911.5299}{{\tt arXiv:0911.5299}}.

\bibitem{Hoche:2010pf}
S.~Hoche, F.~Krauss, M.~Schonherr, and F.~Siegert, {\it {Automating the POWHEG
  method in Sherpa}},  {\em JHEP} {\bf 1104} (2011) 024,
  [\href{http://xxx.lanl.gov/abs/1008.5399}{{\tt arXiv:1008.5399}}].

\bibitem{Alioli:2010xd}
S.~Alioli, P.~Nason, C.~Oleari, and E.~Re, {\it {A general framework for
  implementing NLO calculations in shower Monte Carlo programs: the POWHEG
  BOX}},  {\em JHEP} {\bf 1006} (2010) 043,
  [\href{http://xxx.lanl.gov/abs/1002.2581}{{\tt arXiv:1002.2581}}].

\bibitem{Re:2010jg}
E.~Re, {\it {Single-top production with the POWHEG method}},  {\em PoS} {\bf
  DIS2010} (2010) 172, [\href{http://xxx.lanl.gov/abs/1007.0498}{{\tt
  arXiv:1007.0498}}].

\bibitem{Re:2010bp}
E.~Re, {\it {Single-top Wt-channel production matched with parton showers using
  the POWHEG method}},  {\em Eur.Phys.J.} {\bf C71} (2011) 1547,
  [\href{http://xxx.lanl.gov/abs/1009.2450}{{\tt arXiv:1009.2450}}].

\bibitem{Alioli:2010qp}
S.~Alioli, P.~Nason, C.~Oleari, and E.~Re, {\it {Vector boson plus one jet
  production in POWHEG}},  {\em JHEP} {\bf 1101} (2011) 095,
  [\href{http://xxx.lanl.gov/abs/1009.5594}{{\tt arXiv:1009.5594}}].

\bibitem{Alioli:2010xa}
S.~Alioli, K.~Hamilton, P.~Nason, C.~Oleari, and E.~Re, {\it {Jet pair
  production in POWHEG}},  {\em JHEP} {\bf 1104} (2011) 081,
  [\href{http://xxx.lanl.gov/abs/1012.3380}{{\tt arXiv:1012.3380}}].

\bibitem{Hamilton:2010mb}
K.~Hamilton, {\it {A positive-weight next-to-leading order simulation of weak
  boson pair production}},  {\em JHEP} {\bf 01} (2011) 009,
  [\href{http://xxx.lanl.gov/abs/1009.5391}{{\tt arXiv:1009.5391}}].

\bibitem{Oleari:2010nx}
C.~Oleari, {\it {The POWHEG-BOX}},  {\em Nucl.Phys.Proc.Suppl.} {\bf 205-206}
  (2010) 36--41, [\href{http://xxx.lanl.gov/abs/1007.3893}{{\tt
  arXiv:1007.3893}}].

\bibitem{Oleari:2011ey}
C.~Oleari and L.~Reina, {\it {W b bbar production in POWHEG}},
  \href{http://xxx.lanl.gov/abs/1105.4488}{{\tt arXiv:1105.4488}}.

\bibitem{Kardos:2011qa}
A.~Kardos, C.~Papadopoulos, and Z.~Trocsanyi, {\it {Top quark pair production
  in association with a jet with NLO parton showering}},
  \href{http://xxx.lanl.gov/abs/1101.2672}{{\tt arXiv:1101.2672}}.

\bibitem{Melia:2011gk}
T.~Melia, P.~Nason, R.~Rontsch, and G.~Zanderighi, {\it {$W^+W^+$ plus dijet
  production in the POWHEGBOX}},  {\em Eur. Phys. J.} {\bf C71} (2011) 1670,
  [\href{http://xxx.lanl.gov/abs/1102.4846}{{\tt arXiv:1102.4846}}].

\bibitem{Lavesson:2008ah}
N.~Lavesson and L.~Lonnblad, {\it {Extending CKKW-merging to One-Loop Matrix
  Elements}},  {\em JHEP} {\bf 12} (2008) 070,
  [\href{http://xxx.lanl.gov/abs/0811.2912}{{\tt arXiv:0811.2912}}].

\bibitem{Hamilton:2010wh}
K.~Hamilton and P.~Nason, {\it {Improving NLO-parton shower matched simulations
  with higher order matrix elements}},  {\em JHEP} {\bf 06} (2010) 039,
  [\href{http://xxx.lanl.gov/abs/1004.1764}{{\tt arXiv:1004.1764}}].

\bibitem{Hoche:2010kg}
S.~Hoche, F.~Krauss, M.~Schonherr, and F.~Siegert, {\it {NLO matrix elements
  and truncated showers}},  \href{http://xxx.lanl.gov/abs/1009.1127}{{\tt
  arXiv:1009.1127}}.

\bibitem{Bahr:2008pv}
M.~Bahr {\em et.~al.}, {\it {Herwig++ Physics and Manual}},  {\em Eur. Phys.
  J.} {\bf C58} (2008) 639--707, [\href{http://xxx.lanl.gov/abs/0803.0883}{{\tt
  arXiv:0803.0883}}].

\bibitem{Figy:2003nv}
T.~Figy, C.~Oleari, and D.~Zeppenfeld, {\it {Next-to-leading order jet
  distributions for Higgs boson production via weak-boson fusion}},  {\em Phys.
  Rev.} {\bf D68} (2003) 073005,
  [\href{http://xxx.lanl.gov/abs/hep-ph/0306109}{{\tt hep-ph/0306109}}].

\bibitem{Kleiss:1986re}
R.~Kleiss, {\it {F}rom two to three jets in heavy boson decays: An algorithmic
  approach},  {\em Phys. Lett.} {\bf B180} (1986) 400.

\bibitem{DeCausmaecker:1981bg}
P.~De~Causmaecker, R.~Gastmans, W.~Troost, and T.~T. Wu, {\it {Multiple
  Bremsstrahlung in Gauge Theories at High- Energies. 1. General Formalism for
  Quantum Electrodynamics}},  {\em Nucl. Phys.} {\bf B206} (1982) 53.

\bibitem{Catani:1996vz}
S.~Catani and M.~H. Seymour, {\it {A general algorithm for calculating jet
  cross sections in NLO QCD}},  {\em Nucl. Phys.} {\bf B485} (1997) 291--419,
  [\href{http://xxx.lanl.gov/abs/hep-ph/9605323}{{\tt hep-ph/9605323}}].

\bibitem{Lonnblad:2006pt}
L.~L\mbox{\"{o}}nnblad, {\it {ThePEG, PYTHIA7, Herwig++ and ARIADNE}},  {\em
  Nucl. Instrum. Meth.} {\bf A559} (2006) 246--248.

\bibitem{Sjostrand:2006za}
T.~Sj\mbox{\"{o}}strand, S.~Mrenna, and P.~Skands, {\it {PYTHIA} 6.4 {P}hysics
  and {M}anual},  {\em JHEP} {\bf 05} (2006) 026,
  [\href{http://xxx.lanl.gov/abs/hep-ph/0603175}{{\tt hep-ph/0603175}}].

\bibitem{Gieseke:2003rz}
S.~Gieseke, P.~Stephens, and B.~Webber, {\it {N}ew {F}ormalism for {QCD}
  {P}arton {S}howers},  {\em JHEP} {\bf 12} (2003) 045,
  [\href{http://xxx.lanl.gov/abs/hep-ph/0310083}{{\tt hep-ph/0310083}}].

\bibitem{Chekanov:2003yv}
{\bf ZEUS} Collaboration, S.~Chekanov {\em et.~al.}, {\it {High-$Q^2$ neutral
  current cross sections in $e^+ p$ deep inelastic scattering at
  $\sqrt{s}=318$\,GeV}},  {\em Phys. Rev.} {\bf D70} (2004) 052001,
  [\href{http://xxx.lanl.gov/abs/hep-ex/0401003}{{\tt hep-ex/0401003}}].

\bibitem{Chekanov:2002ej}
{\bf ZEUS} Collaboration, S.~Chekanov {\em et.~al.}, {\it {Measurement of
  high-$Q^2$ $e^-p$ neutral current cross sections at HERA and the extraction
  of $xF_3$}},  {\em Eur. Phys. J.} {\bf C28} (2003) 175,
  [\href{http://xxx.lanl.gov/abs/hep-ex/0208040}{{\tt hep-ex/0208040}}].

\bibitem{Catani:1996gg}
S.~Catani and M.~H. Seymour, {\it {NLO QCD calculations in DIS at HERA based on
  the dipole formalism}},  \href{http://xxx.lanl.gov/abs/hep-ph/9609521}{{\tt
  hep-ph/9609521}}.

\bibitem{Chekanov:2002pv}
{\bf ZEUS} Collaboration, S.~Chekanov {\em et.~al.}, {\it {A ZEUS
  next-to-leading-order QCD analysis of data on deep inelastic scattering}},
  {\em Phys. Rev.} {\bf D67} (2003) 012007,
  [\href{http://xxx.lanl.gov/abs/hep-ex/0208023}{{\tt hep-ex/0208023}}].

\bibitem{Adloff:1999ws}
{\bf H1} Collaboration, C.~Adloff {\em et.~al.}, {\it {Measurements of
  transverse energy flow in deep inelastic-scattering at HERA}},  {\em Eur.
  Phys. J.} {\bf C12} (2000) 595--607,
  [\href{http://xxx.lanl.gov/abs/hep-ex/9907027}{{\tt hep-ex/9907027}}].

\bibitem{Cacciari:2005hq}
M.~Cacciari and G.~P. Salam, {\it {Dispelling the $N^{3}$ myth for the $k_t$
  jet-finder}},  {\em Phys. Lett.} {\bf B641} (2006) 57--61,
  [\href{http://xxx.lanl.gov/abs/hep-ph/0512210}{{\tt hep-ph/0512210}}].

\bibitem{Pumplin:2002vw}
J.~Pumplin {\em et.~al.}, {\it {New generation of parton distributions with
  uncertainties from global QCD analysis}},  {\em JHEP} {\bf 07} (2002) 012,
  [\href{http://xxx.lanl.gov/abs/hep-ph/0201195}{{\tt hep-ph/0201195}}].

\bibitem{Tung:2002vr}
W.-K. Tung, {\it {New generation of parton distributions with uncertainties
  from global QCD analysis}},  {\em Acta Phys. Polon.} {\bf B33} (2002)
  2933--2938, [\href{http://xxx.lanl.gov/abs/hep-ph/0206114}{{\tt
  hep-ph/0206114}}].

\end{thebibliography}\endgroup
\end{document}